\documentclass[twocolumn]{autart}    % Enable this line and disable the 
\usepackage{algorithm}
\usepackage{algorithmicx}
\usepackage{algpseudocode}
\usepackage{graphicx}          % Include this line if your 
\usepackage{mathrsfs}         
\usepackage{amsmath}   
\usepackage{amsfonts}    
\usepackage{amssymb}   
\usepackage{subfigure}
\usepackage{color}
\usepackage{picins}
\usepackage{comment}

\newcommand*{\QEDA}{\hfill\ensuremath{\square}}    

\DeclareMathOperator{\diag}{diag}
\DeclareMathOperator{\Span}{span}
\DeclareMathOperator{\Null}{null}
\DeclareMathOperator{\rank}{rank}
\DeclareMathOperator{\range}{range}
\begin{document}

\begin{frontmatter}

\title{Angle-Based Shape Determination Theory of Planar Graphs with Application to Formation Stabilization (Extended Version)\thanksref{footnoteinfo}} % Title, preferably not more 

\thanks[footnoteinfo]{This work was supported by National Science Foundation of China (Grant Nos. 61751301 and 61533001). Corresponding author: Long Wang. This paper is the extended version of our paper published in Automatica.}

\author[xidian]{Gangshan Jing} \ead{nameisjing@gmail.com},    % Add the 
\author[PolyU]{Guofeng Zhang} \ead{Guofeng.Zhang@polyu.edu.hk},
\author[PolyU]{Heung Wing Joseph Lee} \ead{Joseph.Lee@polyu.edu.hk},
\author[beida]{Long Wang } \ead{longwang@pku.edu.cn}  % (ead) as shown

\address[xidian]{Center for Complex Systems, School of Mechano-Electronic Engineering, Xidian University, Xi'an 710071, China}
\address[PolyU]{Department of Applied Mathematics, Hong Kong Polytechnic University, Hong Kong, China}
\address[beida]{Center for Systems and Control, College of Engineering, Peking University, Beijing 100871, China}       % here.

                             % keyword list or with the 
                                          % help of the Automatica 
                                          % keyword wizard

\begin{abstract}                          % Abstract of not more than 200 words.
This paper presents an angle-based approach for distributed formation shape stabilization of multi-agent systems in the plane. We develop an angle rigidity theory to study whether a planar framework can be determined by angles between segments uniquely up to translations, rotations, scalings and reflections. The proposed angle rigidity theory is applied to the formation stabilization problem, where multiple single-integrator modeled agents cooperatively achieve an angle-constrained formation. During the formation process, the global coordinate system is unknown for each agent and wireless communications between agents are not required. Moreover, by utilizing the advantage of high degrees of freedom, we propose a distributed control law for agents to stabilize a target formation shape with desired orientation and scale. Simulation examples are performed for illustrating effectiveness of the proposed control strategies.
\end{abstract}

\end{frontmatter}

\section{Introduction}
A multi-agent formation stabilization problem is to design a decentralized control law for a group of mobile agents to stabilize a prescribed formation shape. An associated  fundamental problem is: how to determine the geometric shape of a graph embedded in a space, based on some local constraints such as displacements, distances and bearings.

A straightforward approach for determining a shape is constraining the location of each vertex in the graph. A position-based formation strategy usually takes large costs and is unnecessary when the position of each agent is not strictly required. For reduction of information exchange and improvement of robustness of the control strategy, a displacement-constrained formation method, which determines the target formation shape by relative positions between agents, has been extensively studied \cite{Fax04,Ren07,Xiao09,Coogan12,Jing17}. This method is also called consensus-based formation since the formation problem can often be transformed into a consensus problem, which is a hot topic being widely studied \cite{Saber07,Wang07,Wang10,Jing17}. The investigations of displacement-based formation show that the shape of a graph can be determined by inter-agent displacements uniquely if the graph is connected. A disadvantage of displacement-based formation control is the requirement of the global coordinate system.

During the last decade, distance-based shape control gained a lot of attention since it has no requirement of the global coordinate system for each agent \cite{Anderson08,Krick09,Yu09,Summers11,Oh11,Zelazo15,Mou16,Sun17,Chen17}. Different from the displacement-based approach, for a noncomplete graph embedded in a space, it is not straightforward to answer that whether its shape can be determined by edge lengths uniquely. A tool of great utility to deal with this problem is the traditional graph rigidity theory (we will refer to this theory as distance rigidity theory in this paper) \cite{Asimow78,Hendrickson92,Liberti14}, which has been studied intensively in the area of mathematics.

In more recent years, bearing-constrained formation control attracted many interests due to the low costs of bearing measurements \cite{Eren03,Eren12,Zelazo14,Zelazo15,Zelazo151,Bishop11,Zhao16,Schiano16,Michieletto16,Zhao17}. In this issue, the formation shape is constrained by inter-agent bearings. To distinguish what kind of shapes can be uniquely determined by inter-agent bearings, the authors in \cite{Eren12} and \cite{Zhao16} proposed the bearing rigidity theory. Compared to distance-constrained formation control, an advantage of bearing-constrained formation strategy is the fact that no restriction on scale of the target formation is imposed. As a result, it is simple to control the scale of a bearing-based formation, which benefits for obstacle avoidance, see \cite{Zhao17}. Unfortunately, similar to the displacement-based approach, a bearing-constrained formation requires either the global coordinate system for each agent or developing observers based on inter-agent communications, \cite{Zhao16}. In \cite{Zelazo14,Zelazo151,Schiano16,Michieletto16}, the authors achieved bearing-based formation control in the absence of the global coordinate system, but each agent should have a controllable quantity determining the relationship between the local body frame and the global coordinate frame.

Besides the above-mentioned investigations, there are some other issues associated with formation control and formation strategies, for more details, we refer the readers to \cite{Wang04,Oh15,Lin16,Aranda16,Jing18}.

This paper studies the angle-constrained formation problem in the plane, in which the target formation shape is the shape of a planar graph (In this paper, planar graphs refer to graphs in the plane), and will be encoded by angles between pair of edges joining a common vertex. Similar issues have been reported in the literature. In \cite{Eren03}, the authors discussed the possibility of an angle-based formation approach and presented some initial results.  In another relevant reference \cite{Zhao14}, the authors solved the cyclic formation problem by constraining the angle subtended at each vertex by its two neighbors. In this case, the cyclic formation can be stretched while preserving invariance of each angle, thus the target formation cannot be accurately stabilized. In contrast to \cite{Zhao14}, we study how to stabilize a formation shape via angle constraints, such that the stabilized formation is congruent to the target formation up to translations, rotations, scalings and reflections. In \cite{Buckley17}, the authors presented infinitesimally shape-similar motions preserving angles, but they did not give an approach for determining rigidity by angles only.

Our contributions can be summarized as follows. (i). Enlightened by distance rigidity theory and bearing rigidity theory, we propose an angle rigidity theory to study whether the shape of a planar graph can be uniquely determined by angles only; see Section \ref{sec angle rigidity}. (ii). We prove that for a framework in the plane, infinitesimal angle rigidity is equivalent to infinitesimal bearing rigidity; Theorem \ref{th IAR=IBR}. From \cite{Zhao17generic}, infinitesimal angle rigidity is also a generic property of the graph. (iii). We show that for a framework embedded by a triangulated Laman graph, once it is strongly nondegenerate, it can always be determined by angles uniquely up to translations, rotations, scalings and reflections; see Theorem \ref{th a unique shape}. (iv). We propose a distributed control law for achieving formation shape stabilization based on the angle rigidity theory. It is shown that our control strategy can locally exponentially stabilize multiple agents to form an infinitesimally angle rigid formation in the plane; see Theorem \ref{th exponentially stable}. (v). We design a distributed control law, which can steer all agents to form a target formation shape with prescribed orientation and scale; see Theorem \ref{th stability of maneuver control}. Note that in the literature of formation maneuver control \cite{Coogan12,Sun17,Zhao17}, controlling orientation and scale of a formation usually cannot be achieved simultaneously.

The advantages of angle-based formation approach are threefold. (i). Each agent only has to measure relative displacements from neighbors with respect to its local coordinate system. (ii). No wireless communications between agents are required. (iii). Compared to displacement-, distance- and bearing-based approaches, an angle-constrained shape has higher degrees of freedom. More precisely, angles are invariant to motions including translations, rotations and scalings, while inter-agent displacements, distances and bearings are only invariant to a subset of these motions. As a result, it is more convenient to achieve formation maneuver control by using angle constraints. In \cite{Zelazo14,Zelazo151,Schiano16,Michieletto16}, the formation constraints are also invariant to translations, rotations, scalings and reflections. Nevertheless, the trivial rotation in these papers consists of a rotation of the framework in the global coordinate frame, and a rotation of each agent in its local coordinate frame with the same angular speed as that of the whole framework.

The paper is structured as follows. Section \ref{sec preliminaries} introduces some preliminaries of distance- and bearing rigidity theory. Section \ref{sec angle rigidity} presents the angle rigidity theory. Section \ref{sec formation} firstly proposes a distributed control law for achieving formation stabilization based on angle rigidity theory, then proposes a distributed maneuver control law for stabilizing a formation shape with pre-specified orientation and scale. Section \ref{sec simulation} presents an application example to verify validity of the formation strategy. Section \ref{sec conclusion} concludes the whole paper.

Notations: Throughout this paper, $\mathbb{R}$ denotes the set of real numbers; $\mathbb{R}^n$ is the $n-$dimensional Euclidean space; $||\cdot||$ stands for the Euclidean norm; $X^T$ means the transpose of matrix $X$; $\otimes$ is the Kronecker product. $\range(X)$, $\Null(X)$ and $\rank(X)$ denote the range space, null space, and the rank of matrix $X$; $I_n$ represents the $n\times n$ identity matrix; $A\setminus B$ is the set of those elements of $A$ not belonging to $B$; A vector $p=(p_1^T,\cdots,p_s^T)^T$ with $p_i\in\mathbb{R}^2$, $i=1,\cdots,s$ is said to be degenerate if $p_1,\cdots,p_s$ are collinear; $\textrm{O(2)}$ is the orthogonal group in $\mathbb{R}^2$; $\mathscr{R}_o(\theta)=\begin{pmatrix}
\cos\theta & -\sin\theta \\
\sin\theta & ~\cos\theta
\end{pmatrix}$ is the 2-dimensional rotation matrix associated with $\theta\in [0,2\pi)$; $\mathscr{R}_e(\theta)=\mathscr{R}_o(\theta)\bar{I}$ with $\bar{I}=\begin{pmatrix}
1 & 0 \\
0 & -1
\end{pmatrix}$ is the 2-dimensional reflection matrix associated with $\theta\in [0,2\pi)$; $x^{\perp}=\mathscr{R}_o(\frac{\pi}{2})x$ for $x\in\mathbb{R}^2$. For $X_i\in\mathbb{R}^{a\times b}$, $i=1,\cdots,q$, we denote $\diag(X_i)=\textrm{blockdiag}\{X_1,\cdots, X_q\}\in\mathbb{R}^{qa\times qb}$.

An undirected graph with $n$ vertices and $m$ edges is denoted as $\mathcal{G}=(\mathcal{V},\mathcal{E})$, where $\mathcal{V}=\{1,\cdots,n\}$ and $\mathcal{E}\subset\mathcal{V}\times\mathcal{V}$ denote the vertex set and the edge set, respectively. Here we do not distinguish $(i,j)$ and $(j,i)$ in $\mathcal{E}$. The incidence matrix is represented by $H=[h_{ij}]$, which is a matrix with rows and columns indexed by edges and vertices of $\mathcal{G}$ with an orientation. $h_{ij}=1$ if the $i$th edge sinks at vertex $j$, $h_{ij}=-1$ if the $i$th edge leaves vertex $j$, and $h_{ij}=0$ otherwise. It is well-known that $\rank(H)=n-1$ if and only if graph $\mathcal{G}$ is connected.  Let $\mathcal{K}$ denote a complete graph with $n$ vertices.

\section{Preliminaries of graph rigidity theory}\label{sec preliminaries}

In this section, we introduce some preliminaries of distance and bearing rigidity theory in the plane, which are taken from \cite{Asimow78,Hendrickson92,Zhao16}. Distance rigidity theory is to answer whether $p$ can be uniquely determined up to translations, rotations, and reflections, by partial length constraints on edges of $\mathcal{G}$, while bearing rigidity theory is to answer whether $p$ can be uniquely determined up to translations and scalings by partial bearing constraints on edges of $\mathcal{G}$. In what follows, we will introduce these two theories in a unified approach.

We refer to a pair $(\mathcal{G},p)$ as a framework, where $\mathcal{G}$ is a graph and $p=(p_1^T,\cdots,p_n^T)^T\in\mathbb{R}^{2n}$ is called a configuration, $p_i$ is the coordinate of vertex $i$, $i=1,\cdots,n$. To define rigidity of a framework $(\mathcal{G},p)$, a smooth {\it rigidity function} $r_{\mathcal{G}}(\cdot):\mathbb{R}^{2n}\rightarrow\mathbb{R}^s$ should be first given, where $s$ is some positive integer. By the given rigidity function $r_{\mathcal{G}}$, several definitions associated with rigidity can be induced as follows.

A framework $(\mathcal{G},p)$ is said to be {\it rigid} if there exists a neighborhood $U_p$ of $p$ such that $r_{\mathcal{G}}^{-1}(r_{\mathcal{G}}(p))\cap U_p=r_{\mathcal{K}}^{-1}(r_{\mathcal{K}}(p))\cap U_p$. $(\mathcal{G},p)$ is {\it globally rigid} if $r_{\mathcal{G}}^{-1}(r_{\mathcal{G}}(p))=r_{\mathcal{K}}^{-1}(r_{\mathcal{K}}(p))$.

An {\it infinitesimal motion} is an assignment of velocities that guarantees the invariance of $r_{\mathcal{G}}(p)$, i.e.,
\begin{equation}\label{infinitesimal}
\dot{r}_{\mathcal{G}}(p)=\frac{\partial r_{\mathcal{G}}(p)}{\partial p}v=0,
\end{equation}
where $v=(v_1^T,\cdots,v_n^T)^T$, $v_i=\dot{p}_i$ is the velocity of vertex $i$. We say a motion is {\it trivial} if it satisfies equation (\ref{infinitesimal}) for any framework with $n$ vertices. A framework is {\it infinitesimally rigid} if every infinitesimal motion is trivial. Denote the {\it rigidity matrix} $ \frac{\partial r_{\mathcal{G}}(p)}{\partial p}$ by $R(p)$. Then equation (\ref{infinitesimal}) is equivalent to $\dot{r}_{\mathcal{G}}(p)=R(p)\dot{p}=0$. Let $T$ be the dimension of the space formed by all trivial motions, then a framework $(\mathcal{G},p)$ is infinitesimally rigid if and only if $\rank(R(p))=2n-T$.

In the traditional graph rigidity theory, the above-mentioned rigidity function $r_{\mathcal{G}}(\cdot)$ is commonly set by the following {\it distance rigidity function}: 
\begin{equation}\label{drigidfun}
D_{\mathcal{G}}(p)=(\cdots,||e_{ij}(p)||^2,\cdots)^T, ~~(i,j)\in\mathcal{E},
\end{equation}
where $e_{ij}(p)=p_i-p_j$.

In \cite{Eren12,Zhao16}, the authors presented bearing rigidity theory by using the following {\it bearing rigidity function} as the rigidity function $r_{\mathcal{G}}(\cdot)$:
\begin{equation}\label{brigidfun}
B_{\mathcal{G}}(p)=(\cdots,g_{ij}^T(p),\cdots)^T, ~~(i,j)\in\mathcal{E},
\end{equation}
where $g_{ij}(p)=\frac{p_i-p_j}{||p_i-p_j||}$.

For a framework in the plane, there are totally $2$ independent translations, $1$ independent rotation, 1 independent scaling. The trivial motions for a framework determined by distances can only be translations and rotations, thus the dimension of trivial motion space should be $T_D=2+1=3$. The trivial motions for a framework determined by bearings are translations and scalings, accordingly, the dimension of trivial motion space is $T_B=2+1=3$.

%In this paper, for notation simplicity, we use DR to denote distance rigid or distance rigidity. We use IDR to denote infinitesimal distance rigidity or infinitesimally distance rigid, we use IBR

The following two lemmas will be used in our paper.
\begin{lem}\cite{Zhao16}\label{le IBR=IDR}
	A framework in $\mathbb{R}^2$ is infinitesimally bearing rigid if and only if it is infinitesimally distance rigid.
\end{lem}

\begin{lem}\cite{Sun17}\label{le noncollinear}
	If a framework in $\mathbb{R}^2$ is infinitesimally distance rigid, then for any vertex $i$, the relative position vectors $p_i-p_j$, $j\in\mathcal{N}_i$ cannot be all collinear.
\end{lem}

It is worth noting that infinitesimal bearing rigidity implies global bearing rigidity \cite{Zhao16}, whereas infinitesimal distance rigidity cannot induce global distance rigidity.
\section{Angle rigidity}\label{sec angle rigidity}

\begin{figure}
	\centering
	\includegraphics[width=8.4cm]{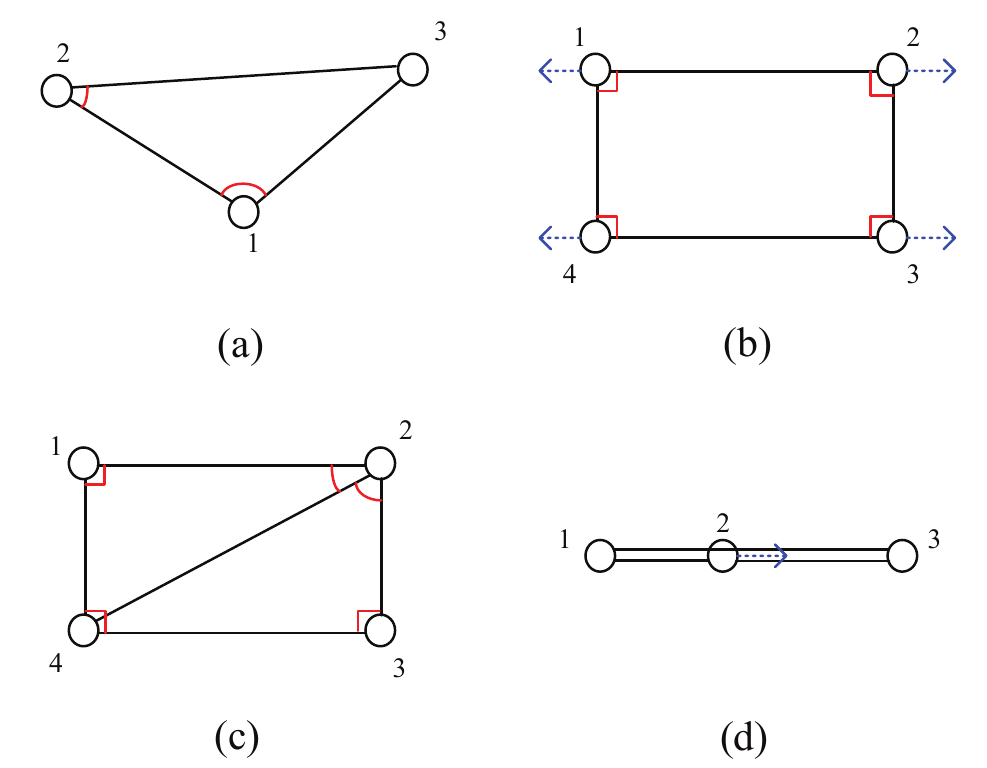}
	\caption{(a) A globally and infinitesimally angle rigid framework with $\mathcal{T}_{\mathcal{G}}^*=\{(1,2,3),(2,1,3)\}$. (b) A framework that is not angle rigid. (c) A globally and infinitesimally angle rigid framework with $\mathcal{T}_{\mathcal{G}}^*=\{(1,2,4),(2,1,4),(2,3,4),(3,2,4),(4,1,3)\}$. (d) A globally angle rigid framework with $\mathcal{T}_{\mathcal{G}}=\{(1,2,3),(2,1,3),(2,1,3)\}$.}
	\label{fig fourframeworks1}
\end{figure}

In this section, we develop an angle rigidity theory to investigate how to encode geometric shapes of graphs embedded in the plane through angles only. For a framework $(\mathcal{G},p)$ in $\mathbb{R}^2$, we will employ $g_{ij}^Tg_{ik}$ as the object we will constrain, which is actually the cosine of the angle between edges $e_{ij}$ and $e_{ik}$. Let $\mathcal{T}_\mathcal{G}=\{(i,j,k)\in\mathcal{V}^3: (i,j),(i,k)\in\mathcal{E}, j<k\}$, then $\{g_{ij}^Tg_{ik}=c_{ijk}:c_{ijk}\in[-1,1],(i,j,k)\in\mathcal{T}_\mathcal{G}\}$ is the set of constraints on all angles in $(\mathcal{G},p)$. We should note that a framework often has redundant angle information for shape determination. For example, in Fig. \ref{fig fourframeworks1} (a), once $g_{12}^Tg_{13}$ and $g_{21}^Tg_{23}$ are available, it holds that $g_{31}^Tg_{32}=\cos(\pi-\arccos(g_{12}^Tg_{13})-\arccos(g_{21}^Tg_{23}))$. That is, the information of partial angles in the graph is often sufficient to recognize a framework. Therefore, by employing a subset $\mathcal{T}_\mathcal{G}^*\subset\mathcal{T}_\mathcal{G}$ with $|\mathcal{T}_\mathcal{G}^*|=w$, we will try to study whether $(\mathcal{G},p)$ can be uniquely determined by  $\{g_{ij}^Tg_{ik}\in[-1,1]:(i,j,k)\in\mathcal{T}_\mathcal{G}^*\}$ based on the angle rigidity theory to be developed in this paper. Note that although $\mathcal{T}_\mathcal{G}^*$ is only a subset of $\mathcal{T}_\mathcal{G}$, the elements in $\mathcal{T}_\mathcal{G}^*$ should involve all vertices in $\mathcal{G}$, otherwise the shape of $(\mathcal{G},p)$ can never be determined. Moreover, we call $\mathcal{T}_\mathcal{G}^*$ the {\it angle index set}.

For a framework $(\mathcal{G},p)$, the {\it angle rigidity function} corresponding to a given angle index set $\mathcal{T}_{\mathcal{G}}^*$ can be written as
\begin{equation}\label{angle rigid function}
f_{\mathcal{T}_\mathcal{G}^*}(p)=(\cdots,g_{ij}^T(p)g_{ik}(p),\cdots)^T, ~~(i,j,k)\in\mathcal{T}_\mathcal{G}^*.
\end{equation}
For the sake of notational simplicity, we denote $f_{\mathcal{G}}(p)=f_{\mathcal{T}_\mathcal{G}}(p)$.

It is easy to see that whether $f_{\mathcal{T}_\mathcal{G}^*}(p)$ can determine a unique shape congruent to $p$ is determined by the choice of $\mathcal{T}_{\mathcal{G}}^*$. As a result, the definitions of angle rigidity must be associated with $\mathcal{T}_{\mathcal{G}}^*$. We present the following definitions.

\begin{defn}\label{ar}
	A framework $(\mathcal{G},p)$ is angle rigid if there exist an angle index set $\mathcal{T}_\mathcal{G}^*$ and a neighborhood $U_p$ of $p$ such that $f_{\mathcal{T}_\mathcal{G}^*}^{-1}(f_{\mathcal{T}_\mathcal{G}^*}(p))\cap U_p=~f_{\mathcal{K}}^{-1}(f_{\mathcal{K}}(p))\cap U_p$.
\end{defn}
\begin{defn}\label{gar}
	A framework $(\mathcal{G},p)$ is globally angle rigid if there exists an angle index set $\mathcal{T}_\mathcal{G}^*$ such that $f_{\mathcal{T}_\mathcal{G}^*}^{-1}(f_{\mathcal{T}_\mathcal{G}^*}(p))=f_{\mathcal{K}}^{-1}(f_{\mathcal{K}}(p))$.
\end{defn}
\begin{defn}
	A framework $(\mathcal{G},p)$ is minimally angle rigid if $(\mathcal{G},p)$ is angle rigid, and deletion of any edge will make $(\mathcal{G},p)$ not angle rigid.
\end{defn}

By these definitions, the frameworks (a) and (c) in Fig. \ref{fig fourframeworks1} are both globally angle rigid. For the framework (b), by moving the vertices along the blue arrows, $f_{\mathcal{G}}$ is invariant but the shape is deformed, thus (b) is not angle rigid. For the framework (d), since the graph is complete, it obviously holds $f_{\mathcal{G}}^{-1}(f_{\mathcal{G}}(p))=f_{\mathcal{K}}^{-1}(f_{\mathcal{K}}(p))$, thus (d) is globally angle rigid. Note that the shape of (d) still cannot be determined by angles uniquely.

Similar to distance and bearing rigidity theory, we define the {\it infinitesimal angle motion} as a motion preserving the invariance of $f_{\mathcal{T}_\mathcal{G}^*}(p)$. The velocity $v=\dot{p}$ corresponding to an infinitesimal motion should satisfy $\dot{f}_{\mathcal{T}_\mathcal{G}^*}(p)=0$, which is equivalent to the following equation
\begin{equation}\label{ginfinitesimal}
\dot{g}_{ij}^Tg_{ik}+g_{ij}^T\dot{g}_{ik}=0,~~(i,j,k)\in\mathcal{T}_\mathcal{G}^*.
\end{equation}
From \cite{Zhao16}, $\frac{\partial g_{ij}}{\partial e_{ij}}=\frac{1}{||e_{ij}||}P_{ij}$, where $P_{ij}\triangleq P(g_{ij})$, $P(\cdot):\mathbb{R}^2\rightarrow\mathbb{R}^{2\times 2}$ is a projection matrix defined as $P(x)= I_2-xx^T$, $x\in\mathbb{R}^2$ is a unit vector. Then we have $\dot{g}_{ij}=\frac{1}{||e_{ij}||}P_{ij}\dot{e}_{ij}$. Let $g(p)=(\cdots,g_{ij}^T(p),\cdots)^T$, where $(i,j)\in\mathcal{E}$, and $R_g\triangleq\frac{\partial f_{\mathcal{T}_\mathcal{G}^*}}{\partial g}$. It follows from the chain rule that
$$\dot{f}_{\mathcal{T}_\mathcal{G}^*}=\frac{\partial f_{\mathcal{T}_\mathcal{G}^*}}{\partial g} \frac{\partial g}{\partial e} \frac{\partial e}{\partial p}\dot{p}=R_g(p) \diag(\frac{P_{ij}}{||e_{ij}||})\bar{H}\dot{p}= R_{\mathcal{T}_{\mathcal{G}}^*}(p)\dot{p},$$
where $\bar{H}=H\otimes I_2$, $R_{\mathcal{T}_{\mathcal{G}}^*}(p)\triangleq R_g(p) \diag(\frac{P_{ij}}{||e_{ij}||})\bar{H}=R_g(p)R_B(p)\in\mathbb{R}^{w\times 2n}$ is termed the {\it angle rigidity matrix}, $R_B=\frac{\partial g(p)}{\partial p}$ is actually the bearing rigidity matrix. Therefore, equation (\ref{ginfinitesimal}) is equivalent to $R_{\mathcal{T}_{\mathcal{G}}^*}(p)\dot{p}=0$.

Next we define infinitesimal angle rigidity, to do this, we should distinguish all trivial motions for an angle-constrained geometric shape. By an intuitive observation, the motions always preserving invariance of angles in the framework are translations, rotations, and scalings. Therefore, the dimension of the trivial motion space is $2+1+1=4$. Note that the trivial motion space is always a subspace of $\Null(R_{\mathcal{T}_{\mathcal{G}}^*})$, implying that $\dim(\Null(R_{\mathcal{T}_{\mathcal{G}}^*}))\geq 4$. We present the following definition.

\begin{defn}\label{de infinitesimal}
	A framework $(\mathcal{G},p)$ is infinitesimally angle rigid if there exists an angle index set $\mathcal{T}_\mathcal{G}^*$ such that every possible motion satisfying (\ref{ginfinitesimal}) is trivial, or equivalently, $\dim(\Null(R_{\mathcal{T}_{\mathcal{G}}^*}))=4$.
\end{defn}

By this definition, the frameworks in Fig. \ref{fig fourframeworks1} (a) and (c) are both infinitesimally angle rigid. The frameworks (b) and (d) are not infinitesimally angle rigid since they both have nontrivial infinitesimal angle motions, which are interpreted by the arrows in blue.

In this paper, we say an angle index set $\mathcal{T}_\mathcal{G}^*$ {\it supports} or {\it is suitable for} (global, minimal, infinitesimal) angle rigidity of $(\mathcal{G},p)$ if $\mathcal{T}_\mathcal{G}^*$ makes the condition in the corresponding definition valid. We say $\mathcal{T}_\mathcal{G}^*$ is {\it minimally suitable} if $\mathcal{T}_\mathcal{G}^*$ is suitable and no proper subset of $\mathcal{T}_\mathcal{G}^*$ can be suitable. It is easy to see from Definitions \ref{ar}-\ref{de infinitesimal} that the angle rigidity property of a framework $(\mathcal{G},p)$ is completely dependent on $\mathcal{G}$ and $p$. After $(\mathcal{G},p)$ is given, whether a suitable $\mathcal{T}_\mathcal{G}^*$ exists becomes certain. However, even for an angle rigid framework, there may exist $\mathcal{T}_\mathcal{G}^*$ such that the conditions in the angle rigidity definitions are invalid. For example, if we choose $\mathcal{T}_\mathcal{G}^*=\mathcal{T}_T$ where $T$ is a spanning tree of $\mathcal{G}$, $\mathcal{T}_\mathcal{G}^*$ can never support angle rigidity of $(\mathcal{G},p)$. On the other hand, there may exist multiple choices of $\mathcal{T}_\mathcal{G}^*$ supporting angle rigidity of a rigid framework. In Subsection \ref{subsec: construction}, Algorithm \ref{alg:T_G^*} will be given to construct a suitable angle index set. 

The following lemma gives the specific form of trivial motions preserving invariance of angles.
\begin{lem}\label{le null(R_a)}
	The trivial motion space for angle rigidity in $\mathbb{R}^2$ is $\mathcal{S}=\mathcal{S}_r\cup\mathcal{S}_s\cup\mathcal{S}_t$, where $\mathcal{S}_r=\{(I_n\otimes \mathscr{R}_o(\frac{\pi}{2}))p\}$ is the space formed by infinitesimal rotations, $\mathcal{S}_s=\Span\{p\}$ is the space formed by infinitesimal scalings, $\mathcal{S}_t=\Null(\bar{H})=\{\mathbf{1}_n\otimes (1,0)^T, \mathbf{1}_n\otimes (0,1)^T\}$ is the space formed by infinitesimal translations.
\end{lem}

\textbf{Proof.} In \cite{Zhao16}, the authors showed that $\mathcal{S}_s$ and $\mathcal{S}_t$ are scaling and translational motion spaces, respectively, and always belong to $\Null(R_B(p))$. Since $R_{\mathcal{T}_{\mathcal{G}}^*}(p)=R_g(p)R_B(p)$, it is straightforward that $\mathcal{S}_s\cup\mathcal{S}_t\subseteq \Null(R_{\mathcal{T}_{\mathcal{G}}^*}(p))$. Next we show $\mathcal{S}_r\subseteq \Null(R_{\mathcal{T}_{\mathcal{G}}^*}(p))$.

Let $\eta^T=\frac{\partial g_{ij}^Tg_{ik}}{\partial g}$ be an arbitrary row of $R_g$. It suffices to show $\eta^TR_B(p)(I_n\otimes \mathscr{R}_o(\frac{\pi}{2}))p=0$. Note that $\eta=(\mathbf{0},g_{ik}^T,\mathbf{0},g_{ij}^T,\mathbf{0})^T$, which follows $\eta^T\diag(\frac{P_{ij}}{||e_{ij}||})=(\mathbf{0},g_{ik}^TP_{ij}/||e_{ij}||,\mathbf{0},g_{ij}^TP_{ik}/||e_{ik}||,\mathbf{0})$. Note also that $\bar{H}(I_n\otimes \mathscr{R}_o(\frac{\pi}{2}))p=(H\otimes I_2)(I_n\otimes \mathscr{R}_o(\frac{\pi}{2}))p=(I_m\otimes \mathscr{R}_o(\frac{\pi}{2}))(H\otimes I_2)p=(I_m\otimes \mathscr{R}_o(\frac{\pi}{2}))e$, where $e=(\cdots,e_{ij}^T,\cdots)^T$, the order of $e_{ij}$ in the vector $e$ is the same as the one of $g_{ij}$ in the vector $g$. It follows that
\[
\begin{split}
\eta^TR_B&(p)(I_n\otimes \mathscr{R}_o(\frac{\pi}{2}))p\\&=\eta^T\diag(\frac{P_{ij}}{||e_{ij}||})\bar{H}(I_n\otimes \mathscr{R}_o(\frac{\pi}{2}))p\\
&=g_{ik}^T\frac{P_{ij}}{||e_{ij}||}\mathscr{R}_o(\frac{\pi}{2})e_{ij}+g_{ij}^T\frac{P_{ik}}{||e_{ik}||}\mathscr{R}_o(\frac{\pi}{2})e_{ik}\\
&=g_{ik}^T(I-g_{ij}g_{ij}^T)\mathscr{R}_o(\frac{\pi}{2})g_{ij}+g_{ij}^T(I-g_{ik}g_{ik}^T)\mathscr{R}_o(\frac{\pi}{2})g_{ik}\\
\end{split}
\]
\[
\begin{split}
&=g_{ik}^T\mathscr{R}_o(\frac{\pi}{2})g_{ij}+g_{ij}^T\mathscr{R}_o(\frac{\pi}{2})g_{ik}\\
&=g_{ik}^T(\mathscr{R}_o(\frac{\pi}{2})+\mathscr{R}^T_o(\frac{\pi}{2}))g_{ij}=0.
\end{split}
\]
This completes the proof.
\QEDA

\begin{lem}
A framework $(\mathcal{G},p)$ is infinitesimally angle rigid if and only if $\Null(R_{\mathcal{T}_{\mathcal{G}}}(p))=\mathcal{S}$.
\end{lem}

In \cite{Asimow78}, the authors showed that the set $D_{\mathcal K}^{-1}(D_{\mathcal K}(p))$, which includes all configurations having inter-distance congruent to $p$, is always a manifold of dimension 3. In fact, since an angle-constrained shape has at least 4 degrees of freedom, $f_{\mathcal{K}}^{-1}(f_{\mathcal{K}}(p))$ is a manifold of dimension 4 when $(\mathcal{K},p)$ is infinitesimally angle rigid (i.e., $p$ is a regular point). See the following theorem.

\begin{thm}\label{th manifold}
Let $\mathscr{S}_p \triangleq \{q\in\mathbb{R}^{2n}: q=c(I_n\otimes \mathscr{R})p+\mathbf{1}_n\otimes\xi, \mathscr{R}\in \textrm{O(2)}, c\in\mathbb{R}\setminus\{0\}, \xi\in\mathbb{R}^2\}$. If $(\mathcal{K},p)$ is infinitesimally angle rigid, then $f_{\mathcal{K}}^{-1}(f_{\mathcal{K}}(p))=\mathscr{S}_p$, and  $\mathscr{S}_p$ is a 4-dimensional manifold.
\end{thm}
The proof will be presented in later subsections.

With the aid of Theorem \ref{th manifold}, we can derive the relationship between infinitesimal angle rigidity and angle rigidity, which is as follows.
\begin{thm}\label{th IAR to AR}
	If $(\mathcal{G},p)$ is infinitesimally angle rigid for $\mathcal{T}_\mathcal{G}^*$, then $(\mathcal{G},p)$ is angle rigid for $\mathcal{T}_\mathcal{G}^*$.
\end{thm}
\textbf{Proof.} By \cite[Proposition 2]{Asimow78} and $\rank \frac{\partial f_{\mathcal{T}_\mathcal{G}^*}}{\partial p}=2n-4$, there is a neighborhood $U$ of $p$, such that $f_{\mathcal{T}_\mathcal{G}^*}^{-1}(f_{\mathcal{T}_\mathcal{G}^*}(p))\cap U$ is a manifold of dimension $4$. From Theorem \ref{th manifold}, $f_{\mathcal{K}}^{-1}(f_{\mathcal{K}}(p))$ is also a 4-dimensional manifold. As a result, $f_{\mathcal{T}_\mathcal{G}^*}^{-1}(f_{\mathcal{T}_\mathcal{G}^*}(p))$ and $f_{\mathcal{K}}^{-1}(f_{\mathcal{K}}(p))$ coincide in $U$, implying that $(\mathcal{G},p)$ is angle rigid.
\QEDA

The converse of Theorem \ref{th IAR to AR} is not true. A typical counter-example is the framework $(\mathcal{K},p)$ with $p$ being a degenerate configuration. In this case, $(\mathcal{K},p)$ is globally angle rigid but not infinitesimally angle rigid.

\subsection{The relation to infinitesimal bearing rigidity} \label{subsec relationship}

In this subsection, we will establish some connections between angle rigidity and bearing rigidity \cite{Zhao16,Zhao17generic}. The following theorem shows the equivalence of infinitesimal angle rigidity and infinitesimal bearing rigidity in the plane, which also implies the feasibility of angle-based approach for determining a framework in the plane.
\begin{thm}\label{th IAR=IBR}
A framework $(\mathcal{G},p)$ is infinitesimally angle rigid if and only if it is infinitesimally bearing rigid.
\end{thm}

\textbf{Proof.} See Appendix. A in Section \ref{app A}.
	\QEDA

\begin{rem}
In \cite{Zhao17generic}, the authors proved that infinitesimal bearing rigidity is a generic property of the graph. That is, if $(\mathcal{G},p)$ is infinitesimally bearing rigid, then $(\mathcal{G},q)$ is infinitesimally bearing rigid for almost all configuration $q$. The underlying approach is showing that a framework embedded by a graph is either infinitesimally bearing rigid or not infinitesimally bearing rigid for all generic configurations\footnote{A configuration $p=(p_1^T,\cdots,p_n^T)^T\in\mathbb{R}^{2n}$ is generic if its $2n$ coordinates are algebraically independent \cite{Jing18}. The set of generic configurations in $\mathbb{R}^{2n}$ is dense.}. From Theorem \ref{th IAR=IBR},  infinitesimal angle rigidity is also a generic property of the graph, thus is primarily determined by the graph, rather than the configuration. In fact, angle rigidity is also a generic property of the graph. To show this, it suffices to show that an angle rigid framework $(\mathcal{G},p^*)$ with a generic configuration $p^*$ is always infinitesimally angle rigid. In \cite[Theorem 3.17]{Jing18}, we have shown that a generic configuration $p^*$ must be a regular point, i.e., $\rank(R_{\mathcal{T}_\mathcal{G}}(p^*))=\max_{p\in\mathbb{R}^{2n}}\rank(R_{\mathcal{T}_\mathcal{G}}(p))\triangleq\kappa$. By \cite[Proposition 2]{Asimow78}, there exists a neighborhood $U$ of $p^*$, such that $f_{\mathcal{G}}^{-1}(f_{\mathcal{G}}(p^*))\cap U$ is a manifold of dimension $2n-\kappa$. From the definition of angle rigidity and Theorem \ref{th manifold}, we know that there exists a neighborhood $U'$ of $p^*$, such that $f_{\mathcal{G}}^{-1}(f_{\mathcal{G}}(p^*))\cap U'$ is a manifold of dimension $4$. By definition of the manifold, we have $2n-\kappa=4$. Then $\kappa=2n-4$. That is, $(\mathcal{G},p^*)$ is infinitesimally angle rigid. Hence, angle rigidity is also a generic property of the graph. By a similar approach, it can be easily obtained that global angle rigidity is also a generic property of the graph.
\end{rem}

\begin{figure}
	\centering
	\subfigure[$(\mathcal{G},p)$]{\includegraphics[width=4cm]{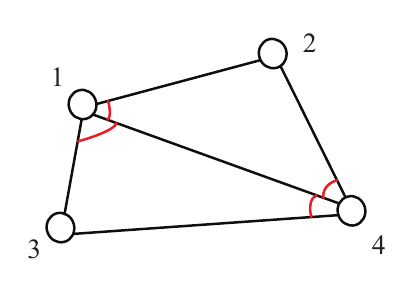}}
	\subfigure[$(\mathcal{G},q)$]{\includegraphics[width=4cm]{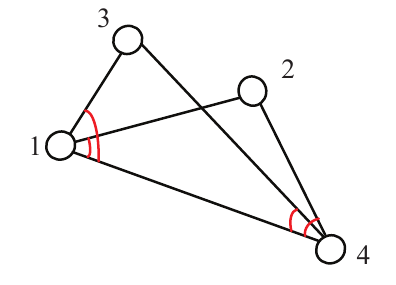}}
	\caption{Both $(\mathcal{G},p)$ and $(\mathcal{G},q)$ are infinitesimally angle rigid for $\mathcal{T}_\mathcal{G}^*=$ $\{(1,2,4), (4,1,2), (1,3,4), (4,1,3)\}$, globally angle rigid for $\bar{\mathcal{T}}_\mathcal{G}^*=\{(1,2,3)\}\cup\mathcal{T}_\mathcal{G}^*$.} \label{fig An IAR framework}
\end{figure}

\begin{rem}
From Definition \ref{de infinitesimal}, we can conclude that the minimal number of angle constraints for achieving infinitesimal angle rigidity is $2n-4$. This fact has also been shown in \cite{Eren03}. On the other hand, it has been shown in \cite{Zhao16} that the minimal number of edges for a framework to be infinitesimally bearing rigid is $2n-3$. By Theorem \ref{th IAR=IBR},  the same is true for infinitesimal angle rigidity.
\end{rem}

Consider a framework $(\mathcal{G},p)$ in the plane. For distance rigidity theory, it is obvious that the shape of $(\mathcal{G},p)$ can be uniquely determined by $D_{\mathcal{G}}(p)$ if $\mathcal{G}=\mathcal{K}$. For bearing rigidity theory, the authors of \cite{Zhao16} showed that $B_{\mathcal{G}}(p)$ uniquely determines a shape if $(\mathcal{G},p)$ is infinitesimally bearing rigid. However, for angle rigidity theory, it cannot be immediately answered that whether the shape can be uniquely determined by angles between edges. This is because angles are only constraints on relationships between those edges joining a common vertex. Even for a complete graph, if $n>3$, there always exist disjoint edges, the angle between each pair of disjoint edges cannot be constrained directly.

In the following theorem, the connection between $f_{\mathcal{K}}^{-1}(f_{\mathcal{K}}(p))$ and $B_{\mathcal{K}}^{-1}(B_{\mathcal{K}}(p))$ is established.

\begin{thm}\label{th angle congruence=bearing congruence} Given configurations $p,q\in\mathbb{R}^{2n}$, $q\in f_\mathcal{K}^{-1}(f_\mathcal{K}(p))$ if and only if $(I_n\otimes\mathscr{R})^{-1}q\in B_{\mathcal{K}}^{-1}(B_{\mathcal{K}}(p))$, where $\mathscr{R}\in \textrm{O(2)}$.
\end{thm}
\textbf{Proof.} See Appendix. B in Section \ref{app B}.
\QEDA

\begin{figure}
	\centering
	\subfigure[$(\mathcal{G},p)$]{\includegraphics[width=4cm]{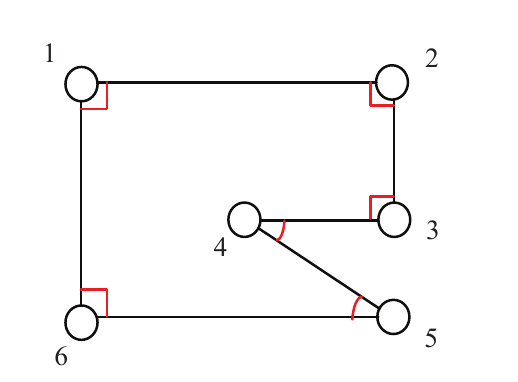}}
	\subfigure[$(\mathcal{G},q)$]{\includegraphics[width=4cm]{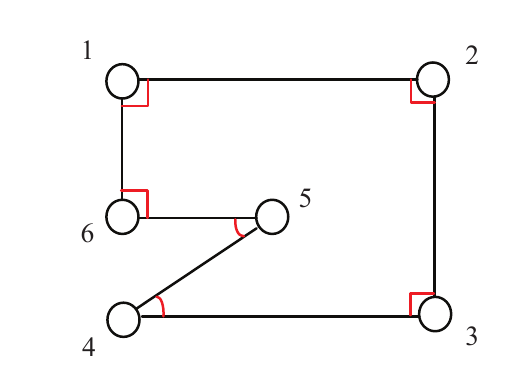}}
	\caption{$f_{\mathcal{G}}(p)=f_{\mathcal{G}}(q)$, but there does not exist $\mathscr{R}\in \textrm{O(2)}$ such that $B_{\mathcal{G}}(p)=(I_m\otimes\mathscr{R}) B_{\mathcal{G}}(q)$. The angles in red are all constrained angles determined by $\mathcal{T}_\mathcal{G}$.}
	\label{fig noncompletegraph}
\end{figure}

\begin{rem}
One can realize that the validity of Theorem \ref{th angle congruence=bearing congruence} will not be lost provided the complete graph $\mathcal{K}$ is replaced with $\mathcal{G}$, where $(\mathcal{G},p)$ is both globally angle rigid and globally bearing rigid. Note that once $\mathcal{K}$ is replaced with a general graph $\mathcal{G}$, Theorem \ref{th angle congruence=bearing congruence} may become invalid. As shown in Fig. \ref{fig noncompletegraph}, although $q\in f_{\mathcal{G}}^{-1}(f_{\mathcal{G}}(p))$, there does not exist $\mathscr{R}\in \textrm{O(2)}$ such that $q\in(I_n\otimes\mathscr{R})^{-1}q\in B_{\mathcal{G}}^{-1}(B_{\mathcal{G}}(p))$.
\end{rem}

It is important to note that Theorem \ref{th angle congruence=bearing congruence} cannot induce equivalence of global angle rigidity and global bearing rigidity. Some examples show that this equivalence holds, but we still have no idea of how to prove it. Nonetheless, we are able to establish the following result.

\begin{thm}
If a framework $(\mathcal{G},p)$ is (globally) angle rigid, then it is (globally) bearing rigid.
\end{thm}
\textbf{Proof.} Suppose $(\mathcal{G},p)$ is angle rigid. Then there exists a neighborhood $U_p$ of $p$ such that $f_{\mathcal{G}}^{-1}(f_{\mathcal{G}}(p))\cap U_p=f_{\mathcal{K}}^{-1}(f_{\mathcal{K}}(p))\cap U_p$. For this $U_p$, consider any $q\in B_{\mathcal{G}}^{-1}(B_{\mathcal{G}}(p))\cap U_p$. It follows from $B_{\mathcal{G}}(p)=B_{\mathcal{G}}(q)$ that $f_{\mathcal{G}}(p)=f_{\mathcal{G}}(q)$. Therefore, $f_{\mathcal{K}}(p)=f_{\mathcal{K}}(q)$. By Theorem \ref{th angle congruence=bearing congruence}, $B_{\mathcal{K}}(p)=(I_m\otimes\mathscr{R})B_{\mathcal{K}}(q)$ for some $\mathscr{R}\in \textrm{O(2)}$. Recall that $B_{\mathcal{G}}(p)=B_{\mathcal{G}}(q)$, we have $\mathscr{R}=I_2$. As a result, $B_{\mathcal{K}}(p)=B_{\mathcal{K}}(q)$, i.e., $q\in B_{\mathcal{K}}^{-1}(B_{\mathcal{K}}(p))$. That is, $(\mathcal{G},p)$ is bearing rigid.

From \cite{Zhao16}, bearing rigidity is equivalent to global bearing rigidity. Since global angle rigidity obviously leads to angle rigidity, it can also induce global bearing rigidity.
\QEDA

To prove Theorem \ref{th manifold}, we introduce the following theorem in \cite{Lee00}.
\begin{thm}\label{th constant-rank}(\cite{Lee00} Constant-Rank Level Set Theorem)
	Let $M$ and $N$ be smooth manifolds, and let $\Phi:M\rightarrow N$ be a smooth map, the Jacobian matrix of $\Phi$ has constant rank $r$. Each level set of $\Phi$ is a properly embedded submanifold of codimension $r$ in $M$.
\end{thm}
\textbf{Proof of Theorem \ref{th manifold}.} From Theorem \ref{th IAR=IBR}, $(\mathcal{K},p)$ is infinitesimally bearing rigid. \cite{Zhao16} shows that $B_{\mathcal{K}}^{-1}(B_{\mathcal{K}}(p))=\{q\in\mathbb{R}^{2n}:q=cp+\mathbf{1}_n\otimes\xi,  c\in\mathbb{R}\setminus\{0\}, \xi\in\mathbb{R}^2\}$. Together with Theorem \ref{th angle congruence=bearing congruence}, there must hold $f_{\mathcal{K}}^{-1}(f_{\mathcal{K}}(p))=\mathscr{S}_p$.

Next we show $\mathscr{S}_p$ is a manifold. For any $q\in f_{\mathcal{K}}^{-1}(f_{\mathcal{K}}(p))$, it is obvious that $q=(I_n\otimes \mathscr{R})(cp+\mathbf{1}_n\otimes\xi)$ for some $\mathscr{R}\in \textrm{O(2)}$, scalar $c$ and vector $\xi\in\mathbb{R}^2$. From the chain rule, we have
\[
\begin{split}
\rank \frac{\partial f_{\mathcal{K}}(q)}{\partial q}&=\rank\frac{\partial f_{\mathcal{K}}(p)}{\partial c(I_n\otimes \mathscr{R})p}\\&=\rank\big[\frac{\partial f_{\mathcal{K}}(p)}{\partial p}\frac{1}{c}(I_n\otimes \mathscr{R}^{-1})\big]\\&=2n-4.
\end{split}
\]
Note that $f_{\mathcal{K}}:\mathbb{R}^{2n}\rightarrow\mathbb{R}^{|\mathcal{T}_{\mathcal{K}}|}$ is a smooth map, according to Theorem \ref{th constant-rank}, $f_{\mathcal{K}}^{-1}(f_{\mathcal{K}}(p))$ is a properly embedded submanifold of dimension $2n-(2n-4)=4$.
\QEDA

\subsection{Construction of $\mathcal{T}_\mathcal{G}^*$ for infinitesimal angle rigidity}\label{subsec: construction}
	
From Definition \ref{de infinitesimal} it is easy to see that $\mathcal{T}_\mathcal{G}$ is always sufficient to determine whether a framework is infinitesimally angle rigid or not. However, the set of angles determined by $\mathcal{T}_\mathcal{G}$ is usually redundant. To reduce computational cost, we give an algorithm to construct a subset $\mathcal{T}_\mathcal{G}^*\subset\mathcal{T}_\mathcal{G}$, which is also sufficient to determine infinitesimal angle rigidity. In the proof for sufficiency of Theorem \ref{th IAR=IBR}, we have presented an approach for constructing a set $\mathcal{T}_\mathcal{G}^*$, and proved that $\mathcal{T}_\mathcal{G}^*$ is a suitable angle index set. Here we give the following algorithm to implement this procedure.
	
\begin{algorithm}
		\renewcommand{\algorithmicrequire}{\textbf{Input:}}
		\renewcommand{\algorithmicensure}{\textbf{Output:}}
		\caption{Finding a Suitable Angle Index Set $\mathcal{T}_\mathcal{G}^*$ for Infinitesimal Angle Rigidity}
		\label{alg:T_G^*}
		\begin{algorithmic}[1]
			\Require An infinitesimally angle rigid framework $(\mathcal{G},p)$ with $p=(p_1^T,\cdots,p_n^T)^T\in\mathbb{R}^{2n}$.
			\Ensure $\mathcal{T}_{\mathcal{G}}^*$
			\State Initialize  $\mathcal{T}_{\mathcal{G}}^*\gets\varnothing$
			\ForAll{$i\in\mathcal{V}$}
			\State Initialize  $^i\mathcal{T}_{\mathcal{G}}^*\gets\varnothing$
			\State Compute the neighbor set of $i$ in $\mathcal{G}$, i.e., $\mathcal{N}_i$ 
			\State Select $j_i$ from $\mathcal{N}_i$ randomly, $\hat{\mathcal{N}}_i\gets\{j_i\}\cup\{k\in\mathcal{N}_i:p_i-p_{j_i}~is~collinear~with~p_i-p_k\}$, $\check{\mathcal{N}}_i\gets\mathcal{N}_i\setminus\hat{\mathcal{N}}_i$
			\State $^i\mathcal{T}_{\mathcal{G}}^*\gets~^i\mathcal{T}_{\mathcal{G}}^*\cup (i,j_i,k)$ for all $k\in\check{\mathcal{N}}_i$. Proceed only if {$|\hat{\mathcal{N}}_i|>1$}
			\State Select $k_i$ from $\check{\mathcal{N}}_i$ randomly
			\ForAll{$j\in\hat{\mathcal{N}}_i\setminus\{j_i\}$}
			\State $^i\mathcal{T}_{\mathcal{G}}^*\gets~^i\mathcal{T}_{\mathcal{G}}^*\cup (i,j,k_i)$ if $j<k_i$, $^i\mathcal{T}_{\mathcal{G}}^*\gets~^i\mathcal{T}_{\mathcal{G}}^*\cup (i,k_i,j)$ otherwise
			\EndFor
			\State $\mathcal{T}_{\mathcal{G}}^*\gets~^i\mathcal{T}_{\mathcal{G}}^*$
			\EndFor
			\State \textbf{return} $\mathcal{T}_{\mathcal{G}}^*$
		\end{algorithmic}
\end{algorithm}

Since each vertex has at most $n-1$ neighbors, it is easy to see that the number of elementary operations performed by Algorithm \ref{alg:T_G^*} is at most $n(n-2)$. Hence the time complexity of Algorithm \ref{alg:T_G^*} is $\mathcal{O}(n^2)$.

\begin{figure}
		\centering
		\includegraphics[width=8.4cm]{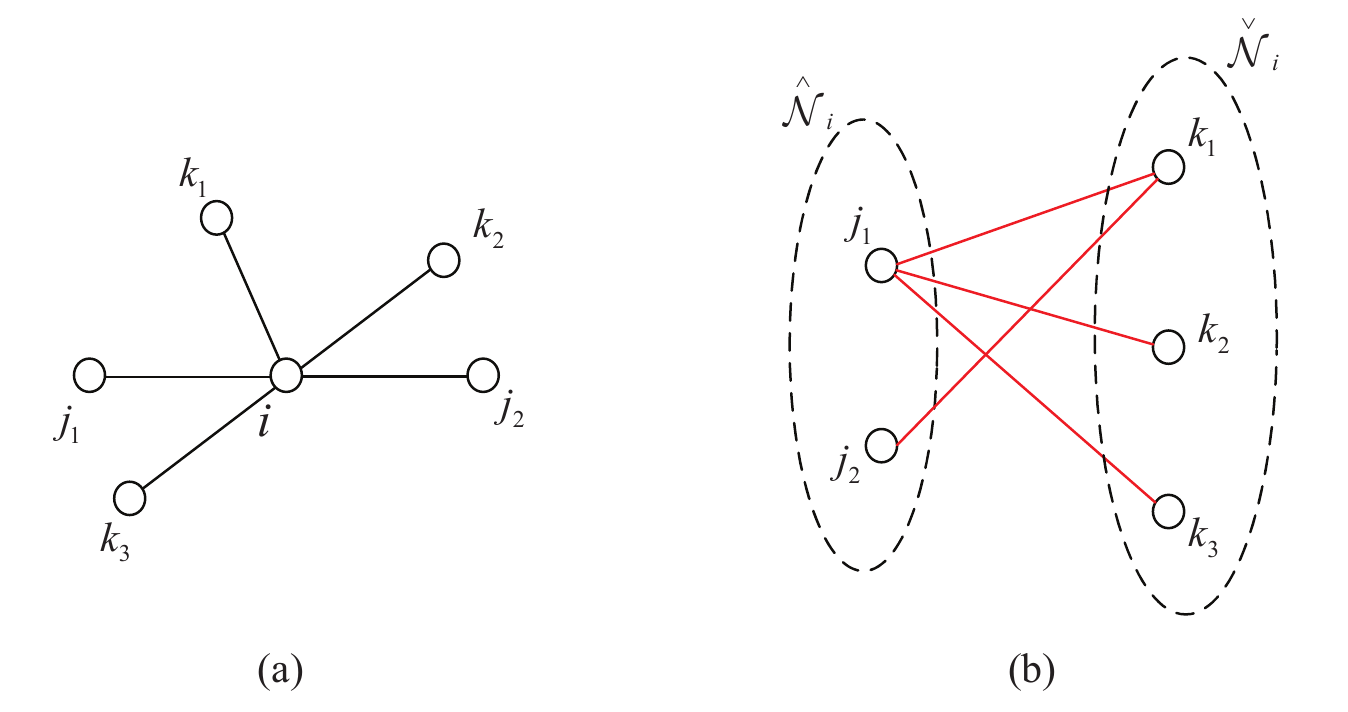}
		\caption{An example to illustrate the construction of $^i\mathcal{T}_{\mathcal{G}}^*$ by Algorithm \ref{alg:T_G^*}. (a) The subgraph composed of vertex $i$ and its neighbors $j_1$, $j_2$, $k_1$, $k_2$, $k_3$. Note that $i,k_2,k_3$ are collinear, $i,j_1,j_2$ are collinear. (b) $\hat{\mathcal{N}}_i=\{j_1,j_2\}$, $\check{\mathcal{N}}_i=\{k_1,k_2,k_3\}$. If $j_s$ and $k_l$ are connected by a red line, it implies that the angle between edge $(i,j_s)$ and edge $(i,k_l)$ is selected to be constrained. This also implies that $(i,j_s,k_l)$ (if $j_s<k_l$) or $(i,k_l,j_s)$ (if $k_l<j_s$) is an element of $^i\mathcal{T}_{\mathcal{G}}^*$.}
		\label{ConstructingTGi}
	\end{figure}

An example of constructing $^i\mathcal{T}_{\mathcal{G}}^*$ by Algorithm \ref{alg:T_G^*} is shown in Fig. \ref{ConstructingTGi}.
	
Note that for an infinitesimally angle rigid framework, the angle index set generated by Algorithm \ref{alg:T_G^*} is suitable but not minimally suitable for infinitesimal angle rigidity. For example, let $(\mathcal{G},p)$ be a minimally angle rigid framework, then $|\mathcal{E}|=2n-3$. For a set $\mathcal{T}_\mathcal{G}^*$ generated by Algorithm \ref{alg:T_G^*}, we have $|\mathcal{T}_\mathcal{G}^*|=\sum_{i\in\mathcal{V}}|^i\mathcal{T}_\mathcal{G}^*|=\sum_{i\in\mathcal{V}}(|\mathcal{N}_i|-1)=2(2n-3)-n=3n-6\geq 2n-4$ for $n\geq2$. Currently we do not have an algorithm to construct a minimally suitable angle index set for an arbitrary infinitesimally angle rigid framework.

	\begin{rem}\label{re IARbuGR}
Although $\mathcal{T}_\mathcal{G}^*$ constructed by Algorithm \ref{alg:T_G^*} supports infinitesimal angle rigidity, it may not support global angle rigidity. As shown in Fig. \ref{fig An IAR framework} (a), by Algorithm \ref{alg:T_G^*}, we can obtain $\mathcal{T}_\mathcal{G}^*=\{(1,2,4), (4,1,2), (1,3,4), (4,1,3)\}$. Although $(\mathcal{G},p)$ is infinitesimally angle rigid, $f_{\mathcal{T}_\mathcal{G}^*}(p)$ may determine an incorrect shape as shown in Fig. \ref{fig An IAR framework} (b). However, if we let $\bar{\mathcal{T}}_\mathcal{G}^*=\{(1,2,3)\}\cup\mathcal{T}_\mathcal{G}^*$, then $f_{\bar{\mathcal{T}}_\mathcal{G}^*}(p)$ can always determine a correct shape. This implies that $(\mathcal{G},p)$ in Fig \ref{fig An IAR framework} (a) is both infinitesimally and globally angle rigid for $\bar{\mathcal{T}}_\mathcal{G}^*(p)$. 
	\end{rem}

In fact, even for a complete graph, it is possible that the geometric shape cannot be determined by angle-only or bearing-only information. A typical example is the degenerate configuration shown in Fig. \ref{fig fourframeworks1} (d). Generally, we hope to determine a framework $(\mathcal{G},p)$ by angles uniquely up to translations, rotations, scalings and reflections in the plane. That is, $f_{\mathcal{G}}^{-1}(f_{\mathcal{G}}(p))=\mathscr{S}_p$. In the next subsection, we will introduce a specific class of frameworks satisfying this condition.
\subsection{A class of frameworks uniquely determined by angles}\label{subsec triangulated laman graphs}

In \cite{Chen17}, the authors introduced a particular class of Laman graphs termed triangulated Laman graphs, which are constructed by a modified Henneberg insertion procedure. In what follows, we will show that the shape of such frameworks can always be determined by angles uniquely. Let $\mathcal{L}_n=(\mathcal{V}_n,\mathcal{E}_n)$ be an $n-$point($n\geq3$) triangulated Laman graph, its definition is as follows.
\begin{defn}\label{de triangulated laman graph}
Let $\mathcal{L}_2$ be the graph with vertex set $\mathcal{V}_2=\{1,2\}$ and edge set $\mathcal{E}_2=\{(1,2)\}$. $\mathcal{L}_n$ ($n\geq3$) is a graph obtained by adding a vertex $n$ and two edges $(n,i)$, $(n,j)$ into graph $\mathcal{L}_{n-1}$ for some $i$ and $j$ satisfying $(i,j)\in\mathcal{E}_{n-1}$.
\end{defn}

Note that the triangulated Laman graph considered here is an undirected graph. Now we give the following result for frameworks embedded by triangulated Laman graphs.

\begin{lem}\label{le strong nondegeneracy to idr}
A triangulated framework $(\mathcal{L}_n,p)$ is infinitesimally distance rigid if and only if $(\mathcal{L}_n,p)$ is strongly nondegenerate, i.e., $p_i$, $p_j$ and $p_k$ are not collinear for any three vertices $i,j,k$ satisfying $(i,j),(j,k),(i,k)\in\mathcal{E}_n$.
\end{lem}

\textbf{Proof.} The proof for sufficiency has been given in \cite{Chen17}. Next we give the proof for necessity. Suppose that strong nondegeneracy does not hold, then there exist $i,j,k\in\mathcal{V}$, such that $(i,j),(j,k),(i,k)\in\mathcal{E}_n$ and $p_i, p_j, p_k$ stay collinear. Note that $(\mathcal{L}_n,p)$ has exactly $2n-3$ edges. Let $R_D(p)=\frac{\partial D_\mathcal{G}(p)}{\partial p}\in\mathbb{R}^{(2n-3)\times 2n}$ be the distance rigidity matrix. To guarantee $\rank(R_D(p))=2n-3$, $R_D(p)$ should be of full row rank. However, it is easy to see that $\frac{\partial ||e_{ij}||^2}{\partial p}$, $\frac{\partial ||e_{ik}||^2}{\partial p}$, and $\frac{\partial ||e_{jk}||^2}{\partial p}$ are always linearly dependent. Hence, $R_D(p)$ cannot be of full row rank, which is a contradiction.
\QEDA

The following theorem shows that the shape of a strongly nondegenerate triangulated framework in the plane can always be uniquely determined by angles.

\begin{thm}\label{th a unique shape}
A triangulated framework $(\mathcal{L}_n,p)$ is strongly nondegenerate 
	
(i) if and only if $(\mathcal{L}_n,p)$ is minimally infinitesimally angle rigid. A minimally suitable angle index set is
\begin{equation}\label{T_L^*}
\mathcal{T}_{\mathcal{L}_n}^*=\{(i,j,k)\in\mathcal{V}^3_n:~~(i,j),(j,k),(i,k)\in\mathcal{E}_n,~i,j<k\};
\end{equation}
	
(ii) only if $(\mathcal{L}_n,p)$ is globally angle rigid. A minimally suitable angle index set is $\mathcal{T}_{\mathcal{L}_n}^\dagger=\mathcal{T}_{\mathcal{L}_n}^*\cup\Delta\mathcal{T}_{\mathcal{L}_n}$, where $\Delta\mathcal{T}_{\mathcal{L}_n}=\{(i,k,l):  k=\min\{\mathcal{N}_i\cap\mathcal{N}_j\cap\mathcal{V}_{l-1}\}, i,j\in\mathcal{N}_l, ~ i<j<l, l=4,\cdots,n\}$ if $n\geq4$, and $\Delta\mathcal{T}_{\mathcal{L}_n}=\varnothing$ otherwise. 
\end{thm}

\textbf{Proof.} See Appendix. C in Section \ref{app C}.
\QEDA

\begin{figure}
\centering
\includegraphics[width=8.4cm]{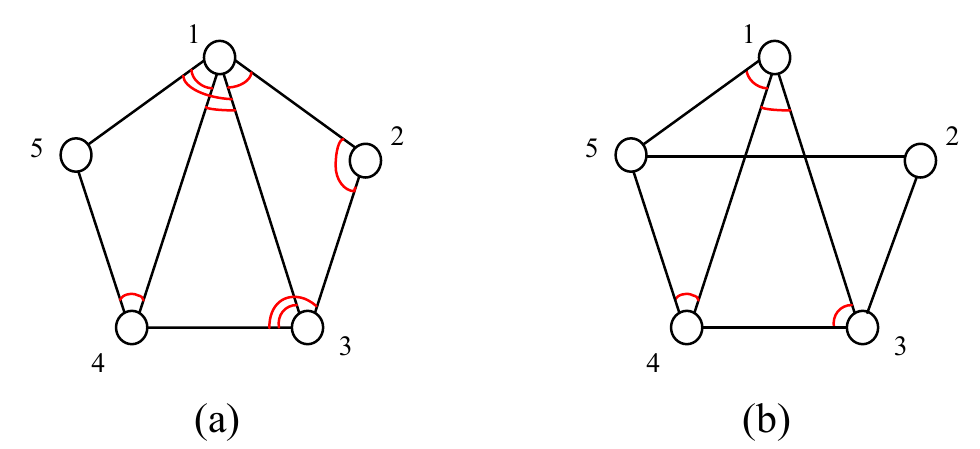}
\caption{(a) A framework embedded by a triangulated Laman graph $\mathcal{L}_5$. $(\mathcal{L}_5,p)$ is infinitesimally angle rigid for $\mathcal{T}_{\mathcal{L}_5}^*=\{(1,2,3),(1,3,4),(1,4,5),(2,1,3),(3,1,4),$ $(4,1,5)\}$, and globally angle rigid for $\mathcal{T}_{\mathcal{L}_5}^\dagger= \mathcal{T}_{\mathcal{L}_5}^*\cup\{(1,2,4), (1,3,5)\}$. The angles in red are constrained angles determined by $\mathcal{T}_{\mathcal{L}_5}^\dagger$. (b) A framework embedded by a Laman graph that is not triangulated. The framework is globally and infinitesimally angle rigid, but $\mathcal{T}_{\mathcal{L}_n}^*$ is not sufficient for its global and infinitesimal angle rigidity. The angles in red are constrained angles determined by $\mathcal{T}_{\mathcal{L}_n}^*$.}
\label{fig two triangulated graphs}
\end{figure}

An example of strongly nondegenerate framework embedded by a triangulated Laman graph is shown in Fig. \ref{fig two triangulated graphs} (a). The angles in red are constrained angles determined by $\mathcal{T}_{\mathcal{L}_5}^\dagger$. The framework in Fig. \ref{fig two triangulated graphs} (b) is both globally angle rigid and infinitesimally angle rigid, but it is not embedded by a triangulated Laman graph. 

It is important to note that strong nondegeneracy is not necessary for a triangulated framework to be globally angle rigid. A simple counterexample is the framework shown in Fig. \ref{fig fourframeworks1} (d). The framework is globally angle rigid, but not strongly nondegenerate. Moreover, the angle index set we give in Theorem \ref{th a unique shape} is only one suitable choice, there are also other choices of the angle index set supporting minimal infinitesimal angle rigidity or global angle rigidity of $(\mathcal{L}_n,p)$.

\section{Application to formation control}\label{sec formation}

In this section, we apply angle rigidity theory to distributed formation control in the plane. The target formation will be characterized by some constraints on angles. In order to form a desired shape, the group of agents are required to meet these constraints via a distributed controller.

\subsection{The formation stabilization problem}

Consider $n$ agents moving in the plane, each agent $i$ has a simple kinematic point dynamics
\begin{equation}\label{dynamics}
\dot{p}_i=u_i, ~~~i\in\mathcal{V},
\end{equation}
where $p_i\in\mathbb{R}^2$ and $u_i\in\mathbb{R}^2$ are the position and control input of agent $i$, respectively, in the global coordinate frame. We consider that the global coordinate system is absent for the agents, each agent $i$ has its own local coordinate system. Let $p^i_j$ be the coordinate of agent $j$'s position with respect to agent $i$'s local coordinate system. Agent $i$ can measure the relative position state $p^i_i-p^i_k$ if $k\in\mathcal{N}_i$.

In this paper, we employ an infinitesimally angle rigid framework $(\mathcal{G},p^*)$ to describe the target formation shape. Each agent is viewed as a vertex of the framework. An interaction link between two agents is regarded as an edge in graph $\mathcal{G}$. That is, $\mathcal{G}$ is also the sensing graph interpreting the interaction relationship between agents.

The target formation shape can be defined as the following manifold:
\begin{multline*}
\mathscr{E}=~\mathscr{S}_{p^*}=\{p\in\mathbb{R}^{2n}: p=c(I_n\otimes \mathscr{R})p^*+\mathbf{1}_n\otimes\xi,\\ \mathscr{R}\in \textrm{O(2)}, c\in\mathbb{R}\setminus\{0\}, \xi\in\mathbb{R}^2\}.
\end{multline*}

For the target formation $(\mathcal{G},p^*)$, we make the following assumption:
\begin{assum}\label{assum1}
	Graph $\mathcal{G}$ contains a triangulated Laman graph $\mathcal{L}_n$ as a subgraph, and $(\mathcal{L}_n,p^*)$ is strongly nondegenerate.
\end{assum}

The set determining all angle constraints is given by
\begin{equation}\label{T_G^f}
\mathcal{T}_\mathcal{G}^F=\{(i,j,k)\in\mathcal{V}^3:(i,j),(j,k),(i,k)\in\mathcal{E},~i,j<k\},
\end{equation}

\begin{rem}\label{re assum1}
Assumption \ref{assum1} is a graphical condition for $(\mathcal{G},p^*)$, and will be the only condition for achieving stability of the target formation. Once Assumption \ref{assum1} holds, it is easy to see that $\mathcal{T}^*_{\mathcal{L}_n}\subset\mathcal{T}_\mathcal{G}^F$, where $\mathcal{T}^*_{\mathcal{L}_n}$ is in form (\ref{T_L^*}). Since we have shown in Theorem \ref{th a unique shape} that $(\mathcal{L}_n,p^*)$ is infinitesimally angle rigid for $\mathcal{T}^*_{\mathcal{L}_n}$, together with $\mathcal{E}_n\subset\mathcal{E}$, it follows that $(\mathcal{G},p^*)$ is infinitesimally angle rigid for $\mathcal{T}_\mathcal{G}^F$. It is also worth noting that strongly nondegenerate configurations form a dense subset of $\mathbb{R}^{2n}$, which is shown in \cite{Chen17}.
\end{rem}

\begin{prob}
Given a set of angle constraints $\mathcal{C}=\{g_{ij}^T(p)g_{ik}(p)=g_{ij}^T(p^*)g_{ik}(p^*)\}$ generated by a framework $(\mathcal{G},p^*)$ satisfying Assumption \ref{assum1}, design a distributed control law for each agent $i$ based on the relative position measurements $\{p^i_i-p^i_j,~j\in\mathcal{N}_i\}$, such that $\mathscr{E}$ is asymptotically stable.
\end{prob}

\subsection{A steepest descent formation controller}

According to the set $\mathcal{C}$, we define the following set as our target equilibrium set of the formation system
\begin{multline}\label{E_F}
\mathscr{E}_F=\{p\in\mathbb{R}^{2n}: g_{ij}^T(p)g_{ik}(p)=g_{ij}^T(p^*)g_{ik}(p^*),\\ (i,j,k)\in\mathcal{T}_\mathcal{G}^F\}.
\end{multline}
Note that $\mathscr{E}$ is a subset of $\mathscr{E}_F$. $\mathscr{E}=\mathscr{E}_F$ if and only if $(\mathcal{G},p^*)$ is globally angle rigid for $\mathcal{T}_\mathcal{G}^F$. In Fig. \ref{fig two triangulated graphs} (a), the framework is only infinitesimally angle rigid, even if all the angle constraints determined by $\mathcal{T}_{\mathcal{G}}^F$ are satisfied, it is possible that the target formation shape is not formed. Nevertheless, from the definition of angle rigidity, for any $q\in\mathscr{E}$, there exists a neighborhood $U$ of $q$, such that $\mathscr{E}\cap U=\mathscr{E}_F\cap U$. Hence, stability of $\mathscr{E}_F$ can still be sufficient for local stability of $\mathscr{E}$. 

Denote $g_{ij}=g_{ij}(p)$, $g_{ij}^*=g_{ij}(p^*)$ for all $(i,j)\in\mathcal{E}$, $\delta_{(i,j,k)}=g_{ij}^Tg_{ik}-g_{ij}^{*T}g_{ik}^*$, $(i,j,k)\in\mathcal{T}_\mathcal{G}^*$. To ensure convergence of (\ref{E_F}), the multi-agent system should minimize the following cost function:
\begin{equation}\label{costfun}
V_F(p)=\frac12\sum_{(i,j,k)\in\mathcal{T}_{\mathcal{G}}^*}(g_{ij}^Tg_{ik}-g_{ij}^{*T}g_{ik}^*)^2 =\frac12\sum_{(i,j,k)\in\mathcal{T}_{\mathcal{G}}^*}\delta_{(i,j,k)}^2 .
\end{equation}
On the basis of function (\ref{costfun}), a gradient-based control strategy can be derived as
\begin{equation}\label{u_i}
\begin{split}
u_i^F&=-\nabla_{p_i}V_F(p)\\&=-\sum_{(j,k)\in\mathcal{N}_{\mathcal{T}_i}}y_1(e_{ij},e_{ik}) -\sum_{(j,k)\in\mathcal{N}_{\mathcal{T}^i}}y_2(e_{ji},e_{jk}), i\in\mathcal{V}
\end{split}
\end{equation}
where $\mathcal{N}_{\mathcal{T}_i}=\{(j,k)\in\mathcal{V}^2:(i,j,k)\in\mathcal{T}_{\mathcal{G}}^F\}$, $\mathcal{N}_{\mathcal{T}^i}=\{(j,k)\in\mathcal{V}^2:(j,i,k)~or~(j,k,i)\in\mathcal{T}_{\mathcal{G}}^F\}$, $y_1(e_{ij},e_{ik})=(g_{ij}^Tg_{ik}-g_{ij}^{*T}g_{ik}^*)(\frac{P_{ij}}{||e_{ij}||}g_{ik}+ \frac{P_{ik}}{||e_{ik}||}g_{ij})$, $y_2(e_{ji},e_{jk})=(g_{ji}^Tg_{jk}-g_{ji}^{*T}g_{jk}^*)\frac{P_{ij}}{||e_{ij}||}g_{kj}$.

Observe that if $(j,i,k)\in\mathcal{T}_{\mathcal{G}}^F$, the control input of agent $i$ includes a term associated with $e_{jk}$. This can be obtained by simple subtraction $e_{ik}-e_{ij}$. From the form of $\mathcal{T}_{\mathcal{G}}^F$ in (\ref{T_G^f}), we have $k,j\in\mathcal{N}_i$. Therefore, $u_i^F$ is a distributed control strategy.

Let $\delta(p)=(\cdots,\delta_{(i,j,k)},\cdots)^T =f_{\mathcal{T}_\mathcal{G}^F}(p)-f_{\mathcal{T}_\mathcal{G}^F}(p^*)$, $(i,j,k)\in\mathcal{T}_{\mathcal{G}}^*$. By the chain rule, the multi-agent system (\ref{dynamics}) with control input (\ref{u_i}) can be written in the following compact form
\begin{equation}\label{compact formation system}
\dot{p}=-\nabla_{p}V_F(p)=-R_{\mathcal{T}_{\mathcal{G}}^F}^T(p)\delta(p).
\end{equation}

The formation system (\ref{compact formation system}) has the following easily checked properties.

\begin{lem}\label{le property}
Under the control law (\ref{u_i}), the following statements hold:
	
(i) The global coordinate system is not required for each agent.
	
(ii) If $p(0)$ is degenerate, then $p(t)=p(0)$ for $t\geq0$.
	
(iii) The centroid $p^o(t)=\frac{1}{n}\sum_{i\in\mathcal{V}}p_i(t)$ and the scale $s(t)=\sqrt{\frac{1}{n}\sum_{i\in\mathcal{V}}||p_i(t)-p^o(t)||^2}$ are both invariant.
\end{lem}
\textbf{Proof.} The proof of (i) is straightforward by a similar approach to those of \cite{Oh11,Sun17}, the validity of (ii) is also easy to verify. Thus their proofs are omitted here. For (iii), observe that $p^o(t)=\frac{1}{n}p^T(t)\mathbf{1}_n$, according to (\ref{compact formation system}), we have 
$$\dot{p}^o=\frac{1}{n}\dot{p}^T\mathbf{1}_n=\frac{1}{n}\delta^TR_{\mathcal{T}_{\mathcal{G}}^F}\mathbf{1}_n=0.$$
To show $\dot{s}=0$, we first note that $\dot{p}^Tp=\delta^TR_{\mathcal{T}_{\mathcal{G}}^F}p=\delta^TR_g\diag(\frac{P_{ij}}{||e_{ij}||})\bar{H}p=0$. It follows that
\[
\begin{split}
\dot{s}&=\frac{2}{\sqrt{n}}\dot{p}^T(p-\mathbf{1}_n\otimes p^o)\\&=-\frac{2}{\sqrt{n}}\delta^T\diag(\frac{P_{ij}}{||e_{ij}||})\bar{H}(\mathbf{1}_n\otimes p^o)=0.
\end{split}
\]
\QEDA

\subsection{Stability analysis}

\begin{thm}\label{th exponentially stable}
For a group of $n\geq 3$ agents with dynamics (\ref{dynamics}) and controller (\ref{u_i}) moving in the plane. Under Assumption \ref{assum1}, for any $q\in\mathscr{E}$, there is a neighborhood $U_q$ of $q$, such that if $p(0)\in U_q$, then $\lim_{t\rightarrow\infty}p(t)=p^\dagger$ for some $p^\dagger\in\mathscr{E}$.
\end{thm}

\textbf{Proof.} For any $q\in\mathscr{E}$, let $\rho=p-q$, $f(p)=-R_{\mathcal{T}_{\mathcal{G}}^F}^T(p)\delta(p)$, expanding $f(p)$ in Taylor series about $q$, we have $f(p)=f(q)+\frac{\partial f(q)}{\partial p}\rho+g(\rho)$. Then (\ref{compact formation system}) is equivalent to 
\begin{equation}\label{rho dynamics}
\dot{\rho}=\frac{\partial f(q)}{\partial p}\rho+g(\rho)=J_f(q)\rho+g(\rho),
\end{equation}
where $J_f(q)=\frac{\partial f(q)}{\partial p}=-\frac{\partial R_{\mathcal{T}_{\mathcal{G}}^F}^T(q)}{\partial p}\delta(q)-R_{\mathcal{T}_{\mathcal{G}}^F}^T(q)\frac{\partial\delta(q)}{\partial p}=-R_{\mathcal{T}_{\mathcal{G}}^F}^T(q)R_{\mathcal{T}_{\mathcal{G}}^F}(q)$. From Lemma \ref{le strong nondegeneracy to idr}, the validity of Assumption 1 implies that $(\mathcal{G},q)$ is infinitesimally angle rigid. Therefore, $J_f(q)$ has 4 zero eigenvalues and the rest are negative real numbers. There must exist an orthonormal transformation $Q\in\mathbb{R}^{2n\times 2n}$ such that $QJ_f(q)Q^T=\diag(\mathbf{0}_{4\times4},\tilde{J})$, where $\tilde{J}$ is Hurwitz. Then (\ref{rho dynamics}) is equivalent to
\begin{equation}\label{pp dynamics}
\begin{split}
&\dot{\varphi}=g_1(\varphi,\psi),\\
&\dot{\psi}=\tilde{J}\psi+g_2(\varphi,\psi),
\end{split}
\end{equation}
where $(\varphi^T,\psi^T)^T=Q\rho$, $(g_1^T,g_2^T)^T=Qg(\rho)$. Note that $\rho=0$ is an equilibrium point of (\ref{rho dynamics}), hence, $g_1(0,0)=0$ and $g_2(0,0)=0$. Since $g(\rho)=f(p)-J_f(q)\rho$, we have $J_g|_{\rho=0}=0$. It follows that $J_{g_1}(0,0)=0$ and $J_{g_2}(0,0)=0$. Observe that $\mathcal{M}=\{(\varphi^T,\psi^T)^T: (\varphi^T,\psi^T)^T=Q\rho, \rho+q\in\mathscr{E}\}$ is a 4-dimensional manifold. We next show $\mathcal{M}$ is a center manifold. Note that $\mathcal{M}$ is invariant since $\mathscr{E}\subset\mathscr{E}_F\subset\{p\in\mathbb{R}^{2n}:f(p)=0\}$. Any equilibria must satisfy $\tilde{J}\psi+g_2(\varphi,\psi)=0$, by implicit function theorem, there is a neighborhood $U$ of the origin, such that $\psi=h(\varphi)$ for $\psi\in U$, where $h(\cdot):\mathbb{R}^4\rightarrow\mathbb{R}^{2n-4}$ is smooth and $h(0)=0$, $J_h(0)=0$. Since $\mathcal{M}$ is a 4-dimensional manifold, there must exist an open set in $\mathbb{R}^4$ that is diffeomorphic to a neighborhood $U'$ of origin in $\mathcal{M}$. Because $\mathcal{M}$ can be represented by $(\varphi^T,h^T(\varphi))^T$ in $U\cap U'$, we conclude that $\mathcal{M}$ is a center manifold. The flow on the manifold $\mathcal{M}$ is governed by the 4-dimensional system $\dot{\xi}=g_1(\xi,h(\xi))$ for sufficiently small $\xi$. Recall that $\mathcal{M}$ is a manifold of equilibria, we have $\dot{\xi}=0$. By center manifold theory \cite{Carr12}, for any sufficiently small $\varphi(0),\psi(0)$, we have $\varphi(t)=\xi(0)+O(e^{-\gamma t})$, $\psi=h(\xi(0))+O(e^{-\gamma t})$ for some $\gamma>0$. This implies that $\lim_{t\rightarrow\infty}(\varphi^T,\psi^T)^T=(\xi^T(0),h^T(\xi(0)))^T\in\mathcal{M}$. It follows that $\lim_{t\rightarrow\infty}p(t)=\lim_{t\rightarrow\infty}Q^T(\varphi^T,\psi^T)^T+q =Q^T(\xi^T(0),h^T(\xi(0)))^T+q=p^\dagger\in\mathscr{E}$. The proof is completed.
\QEDA

The difficulties in achieving global stability of the desired formation shape are two folds: (i). Observe that the equilibrium set of system (\ref{compact formation system}) is $E=\{p\in\mathbb{R}^{2n}:R_{\mathcal{T}_\mathcal{G}^F}^T(p)\delta=0\}$. If $R_{\mathcal{T}_\mathcal{G}^F}\in\mathbb{R}^{|\mathcal{T}_\mathcal{G}^F|\times2n}$ is of full row rank, then $p\in E$ implies $\delta=0$, which yields $p\in\mathscr{E}_F$. However, $R_{\mathcal{T}_\mathcal{G}^F}(p)$ varies as the formation system evolves, it is difficult to determine its rank. Moreover, once $|\mathcal{T}_\mathcal{G}^F|>2n-4$, $R_{\mathcal{T}_\mathcal{G}^F}$ can never be of full row rank. As a result, undesired equilibria often exist for system (\ref{compact formation system}). (ii). Different from displacement, distance and bearing constraints on pairwise agents, each angle constraint involves three agents. The form of $\mathcal{T}_\mathcal{G}^F$ implies that only those angles in triangles can be used as constraints to determine the shape. Nevertheless, in some cases these constraints cannot uniquely determine the desired formation shape. For example, consider the framework in Fig. \ref{fig two triangulated graphs} (a), although its shape can be uniquely determined by angles, it cannot be uniquely determined by angles corresponding to $\mathcal{T}_\mathcal{G}^F$. Fig. \ref{fig formation1} shows a counterexample that under some initial condition, the agents exponentially form an incorrect formation.

Theorem \ref{th exponentially stable} actually means that by implementing the control law (\ref{u_i}), the agents can cooperatively restore the desired formation shape under a small perturbation from any $q\in\mathscr{E}$, and the convergence rate is as fast as $e^{-\gamma t}$ for some $\gamma>0$ dependent on $q$. However, it is uncertain that whether there exists a uniform exponent $\gamma$ for all $q\in\mathscr{E}$. This is because $\mathscr{E}$ is not compact, there does not exist a finite subcover containing $\mathscr{E}$.

\subsection{Orientation and scaling control}\label{subsec maneuver}

We have shown that the angle-constrained formation has 4 degrees of freedom, which is higher than that of displacement-, distance-, and bearing-based formations. This ensures that one advantage of the angle-based formation approach is the convenience of orientation and scaling control. In this subsection, we propose an angle-based control scheme to steer all agents to form a target formation shape with pre-specified orientation and scale. 

Given $(\mathcal{G},p^*)$ as the target formation shape satisfying Assumption \ref{assum1}, a configuration forming the target formation with desired orientation and scale can be written by $p=c^*(I_n\otimes \mathscr{R}_o(\theta^*))p^*+\mathbf{1}_n\otimes\xi$ for some constant $\theta^*\in[0,2\pi)$, $c^*\in\mathbb{R}\setminus\{0\}$ and an arbitrary translational vector $\xi\in\mathbb{R}^2$. It is worth noting that $p_i^*$ here denotes the position of agent $i$ in the global coordinate frame. Let $\tilde{p}=c^*(I_n\otimes\mathscr{R}_o(\theta^*))p^*$, then the target equilibrium can be described as
\begin{equation}
\mathscr{E}_M=\{p\in\mathbb{R}^{2n}: p=\tilde{p}+\mathbf{1}_n\otimes\xi, \xi\in\mathbb{R}^2\}.
\end{equation}

To control the orientation of the formation, it is obviously necessary that some agents should have access to the global coordinate system. To keep the target shape in a precise orientation, we will try to constrain the displacement between two adjacent agents, which is similar to \cite{Sun17}. Since orientation and scale of the ultimate formation are determined by these two agents, we call them leaders. It is noteworthy that any two adjacent agents can be selected as leaders, and controlling their relative position is sufficient to control the orientation and scale of the formation (this fact will be shown later). Moreover, different from \cite{Sun17}, using angle-based approach, the target displacement between leaders can be artificially specified and does not have to satisfy a fixed length constraint. 

Suppose agents $l_1$ and $l_2$ are leaders, $l_1,l_2\in\mathcal{V}$. Then $\tilde{p}_{l_1}-\tilde{p}_{l_2}$ is the displacement of $l_1$ and $l_2$ in the formation with target orientation and scale. Now we summarize the problem that we will deal with in this subsection as below.
\begin{prob}\label{maneuver problem}
	Given a realizable target formation $(\mathcal{G},p^*)$ satisfying Assumption \ref{assum1}, and the target displacement $\tilde{p}_{l_1}-\tilde{p}_{l_2}$ known by agents $l_1$ and $l_2$, design a distributed control law for each agent $i$ based on the relative position measurements $\{p^i_i-p^i_j,~j\in\mathcal{N}_i\}$, such that $\mathscr{E}_M$ is asymptotically stable.
\end{prob}

To solve Problem \ref{maneuver problem}, we consider the following set containing the target equilibrium $\mathscr{E}_M$
\begin{equation}
\mathscr{E}_l=\{p\in\mathscr{E}_F: p_{l_1}-p_{l_2}=\tilde{p}_{l_1}-\tilde{p}_{l_2}\},
\end{equation}
where $\mathcal{E}_F$ is in the form (\ref{E_F}).

The following lemma shows that once $(\mathcal{G},p^*)$ is infinitesimally angle rigid, $\mathscr{E}_l$ and $\mathscr{E}_M$ coincide near each point in $\mathscr{E}_M$. 

\begin{lem}\label{le E_l=E_M}
If $(\mathcal{G},p^*)$ is infinitesimally angle rigid, then for any $q\in\mathscr{E}_M$, there exists a neighborhood $U_q$ of $q$, such that $\mathscr{E}_M\cap U_q=\mathscr{E}_l\cap U_q$.
\end{lem}
\textbf{Proof.} Let $f_l(p)=\begin{pmatrix}
f(p) \\ p_{l_1}-p_{l_2}
\end{pmatrix}\in\mathbb{R}^{|\mathcal{T}_{\mathcal{G}}^F|+2}$, $f_M(p)=(\cdots,(p_i-p_j)^T,\cdots)^T\in\mathbb{R}^{2m}$, it follows that $\mathscr{E}_l=f_l^{-1}(f_l(\tilde{p}))$, $\mathscr{E}_M=f_M^{-1}(f_M(\tilde{p}))$. Since $\mathcal{G}$ must be connected, we have $\rank(\frac{\partial f_M}{\partial p})=\rank(\bar{H})=2n-2$, here $\bar{H}$ is the incidence matrix, according to Theorem \ref{th constant-rank}, $\mathscr{E}_M$ is a 2-dimensional manifold. 

Next we show $\mathscr{E}_l$ is also a 2-dimensional manifold near each $q\in\mathscr{E}_M$. Without loss of generality, suppose $p_{l_1}-p_{l_2}$ is consisted of the $(2k-1)$-th row and $2k$-th row of $\bar{H}p$. Let $S=[S_{ij}]\in\mathbb{R}^{2\times 2m}$ be a matrix with $S_{1,2k-1}=S_{2,2k}=1$, and $S_{ij}=0$ for other $i,j$. Then $S\bar{H}p=p_{l_1}-p_{l_2}$ and $\frac{\partial (p_{l_1}-p_{l_2})}{\partial p}=S\bar{H}$. For any $q\in\mathscr{E}_M$, denote $R_l(q)=\frac{\partial f_l}{\partial p}|_{p=q}=(R_{\mathcal{T}_{\mathcal{G}}^F}^T(q), (S\bar{H})^T)^T$, it is easy to obtain $\Null(R_l(q))=\Null(R_{\mathcal{T}_{\mathcal{G}}^F}(q))\cap \Null(S\bar{H})$. We first notice that $(\mathcal{G},q)$ must be infinitesimally angle rigid, implying $\Null(R_{\mathcal{T}_{\mathcal{G}}^F}(q))=\mathcal{S}$, where $\mathcal{S}$ is the trivial motion space shown in Lemma \ref{le null(R_a)}. We also note that $\Null(S\bar{H})=(\Null(S)\cap \range(\bar{H}))\cup \Null(\bar{H})$. It can be verified that $\Null(R_l(q))=\mathcal{S}\cap \Null(S\bar{H}) =\Null(\bar{H}) =\Span\{\mathbf{1}_n\otimes I_2\}$. Then we obtain $\rank(\frac{\partial f_l}{\partial p}|_{p=q})=2n-2=\max\{\rank(\frac{\partial f_l}{\partial p}): p\in\mathbb{R}^{2n}\}$, i.e., $q$ is a regular point of $f_l$. From \cite[Proposition 2]{Asimow78}, there exists a neighborhood $U$ of $q$, such that $\mathscr{E}_l\cap U$ is a 2-dimensional manifold. Together with $\mathscr{E}_M\subset\mathscr{E}_l$, we have $\mathscr{E}_M\cap U\subset\mathscr{E}_l\cap U$. It follows that $\mathscr{E}_l\cap U_q=\mathscr{E}_M\cap U_q$ for some $U_q\subset U$.
\QEDA

By virtue of Lemma \ref{le E_l=E_M}, when the initial positions of agents are close to $\mathscr{E}_M$, to drive the agents into $\mathscr{E}_M$, it suffices to constrain $p_{l_1}-p_{l_2}$ to be $\tilde{p}_{l_1}-\tilde{p}_{l_2}$ while steering the agents to meet angle constraints determined by $\mathcal{T}_\mathcal{G}^F$. Therefore, we wish the agents to cooperatively minimize the following cost function
\begin{equation}\label{maneuver costfun}
V=V_F+V_M,
\end{equation}
where $V_F$ is in form (\ref{costfun}), $V_M=\frac12||\tilde{p}_{l_1}-\tilde{p}_{l_2}-(p_{l_1}-p_{l_2})||^2$.

We propose the following gradient-based control law 
\begin{equation}\label{u_i^M}
u_i=u_i^F+u_i^M=-\nabla_{p_i}V_F-\nabla_{p_i}V_M, ~~~i\in\mathcal{V}
\end{equation}
where $u_i^F$ in form (\ref{u_i}) is to drive agents to maintain the target shape, $u_i^M$ is for controlling formation orientation and scale.  

It is easy to see that the control law (\ref{u_i^M}) is distributed and $u_i^M=0$ for $i\in\mathcal{V}\setminus\{l_1,l_2\}$. Under (\ref{u_i^M}), property (i) in Lemma \ref{le property} also holds for the formation system, while (ii) in Lemma \ref{le property} becomes invalid. Moreover, during the evolution, the centroid is still invariant, but the formation scale may be changed.

Define the graph $\mathcal{G}_l=(\mathcal{V},\mathcal{E}_l)$, where $\mathcal{E}_l=\{(p_{l_1},p_{l_2})\}$ (we do not distinguish $(p_{l_1},p_{l_2})$ and $(p_{l_2},p_{l_1})$). Let $H_l\in\mathbb{R}^{1\times n}$ be the incidence matrix and $L_l=H_l^TH_l$ be the Laplacian matrix, corresponding to graph $\mathcal{G}_l$. Denote $\bar{p}=p-\tilde{p}$, by using control law (\ref{u_i^M}), the formation system can be written in the following compact form
\begin{equation}\label{maneuver system}
\dot{p}=h_M(p)=-R_{\mathcal{T}_{\mathcal{G}}^F}^T(p)\delta(p)-(L_l\otimes I_2)\bar{p}(p).
\end{equation}
The Jacobian matrix of $h_M$ at the desired equilibrium $\tilde{p}\in\mathscr{E}_l$ is
\[
\begin{split}
J_{h_M}(p)|_{p=\tilde{p}}&=-\frac{\partial R_{\mathcal{T}_{\mathcal{G}}^F}^T(p)}{\partial p}\delta(p)|_{p=\tilde{p}}\\ &~~~~-R_{\mathcal{T}_{\mathcal{G}}^F}(p)\frac{\partial \delta(p)}{\partial p}|_{p=\tilde{p}}- (L_l\otimes I_2)|_{p=\tilde{p}}\\
&=-(R_{\mathcal{T}_{\mathcal{G}}^F}^T(\tilde{p})R_{\mathcal{T}_{\mathcal{G}}^F}(\tilde{p})+L_l\otimes I_2)\triangleq J_M.
\end{split}
\]

The following theorem shows the effectiveness of our control strategy.

\begin{thm}\label{th stability of maneuver control}
For a group of $n\geq 3$ agents with dynamics (\ref{dynamics}) and controller (\ref{u_i^M}) moving in the plane. Under Assumption \ref{assum1}, $\mathscr{E}_M$ is locally exponentially stable.
\end{thm}
\textbf{Proof.} By Lemma \ref{le E_l=E_M}, we only have to show local exponential stability of $\mathscr{E}_l$. Note that system (\ref{maneuver system}) has a similar form to \cite[Equation (9)]{Sun17}. Moreover, $\Null(J_M)=\Null(R_l(\tilde{p}))=\Span\{\mathbf{1}_n\otimes I_2\}$, where $R_l$ is the matrix defined in the proof of Lemma \ref{le E_l=E_M}. Through a process similar to the proof of \cite[Theorem 3]{Sun17}, it can be shown that $\mathscr{E}_l$ is locally exponentially stable. By Lemma \ref{le E_l=E_M}, $\mathscr{E}_M$ is also locally exponentially stable.
\QEDA

\section{Simulations}\label{sec simulation}

In this section, by considering 5 autonomous agents moving in the plane, we present three numerical examples to illustrate the effectiveness of the theoretical findings.

\begin{figure}
\centering
\includegraphics[width=8.5cm]{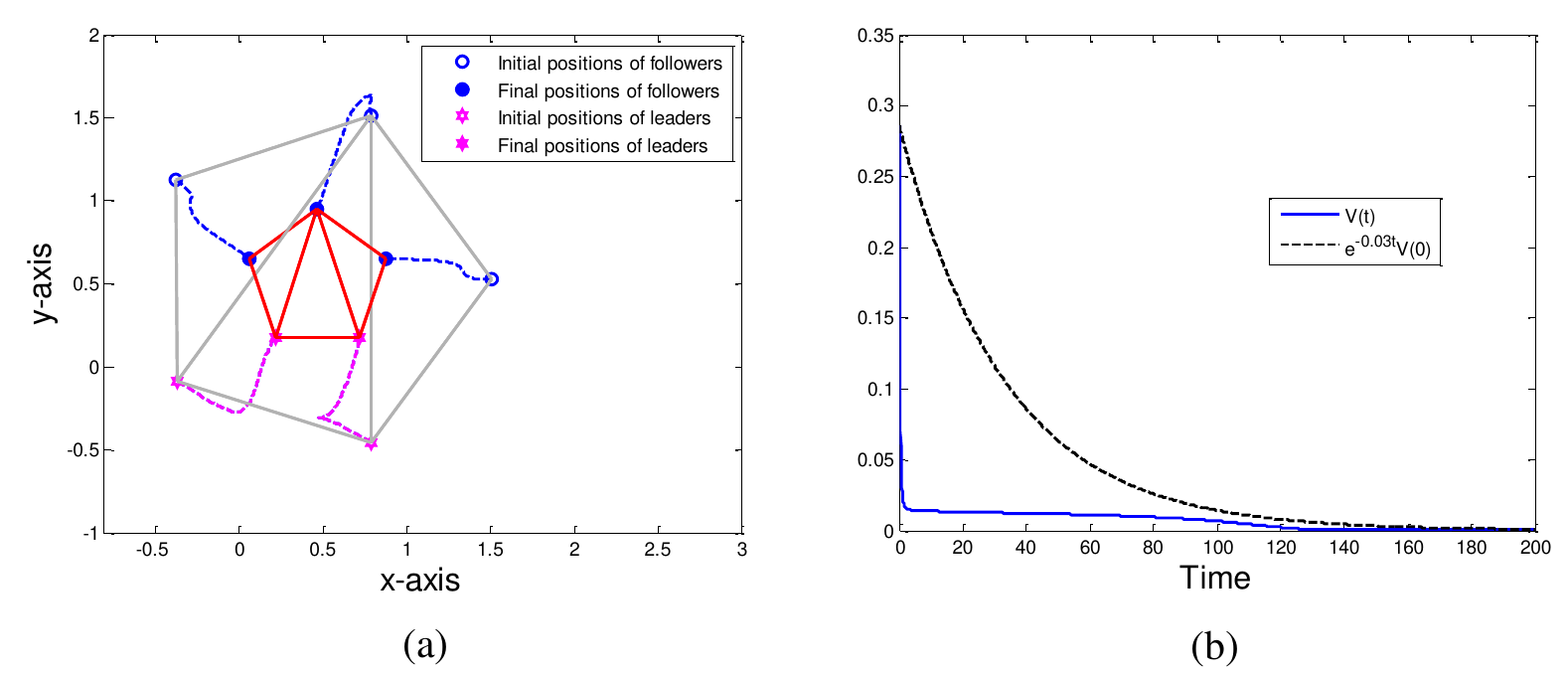}\\
\caption{(a) Under control law (\ref{u_i^M}), the regular pentagon formed by all agents is asymptotically transformed into another regular pentagon with desired orientation and scale. (b) $V(t)$ vanishes to zero exponentially.}\label{fig maneuver}
\end{figure}

\begin{exmp}\label{example1}
Consider regular pentagon described by the framework in Fig. \ref{fig two triangulated graphs} (a) as the target formation shape $(\mathcal{G},p^*)$. The set of desired angle information should be $\{g_{12}^{*T}g_{13}^*=0.8090, g_{13}^{*T}g_{14}^*=0.8090, g_{14}^{*T}g_{15}^*=0.8090, g_{21}^{*T}g_{23}^*=-0.3090, g_{31}^{*T}g_{34}^*=0.3090, g_{41}^{*T}g_{45}^*=0.8090\}$. Note that $\mathcal{G}$ is a triangulated Laman graph, and $(\mathcal{G},p^*)$ is strongly nondegenerate. That is, Assumption \ref{assum1} holds. Without loss of generality, choose $q_i=(\cos(\frac{2\pi}{5}i),\sin(\frac{2\pi}{5}i))^T$, $i=1,\cdots,5$. Then $q=(q_1^T,\cdots,q_n^T)^T\in\mathscr{E}$. Set the initial position vector of the agents as $p(0)=q+r$, where $r\in\mathbb{R}^{10}$ is a perturbation, each component of $r$ is a pseudorandom value drawn from the uniform distribution on $(-0.5,0.5)$. By implementing the control law (\ref{u_i}), Fig. \ref{fig formation} (a) is obtained, which shows that the desired formation shape can be formed by our formation strategy. Fig. \ref{fig formation} (b) describes the evolution of $V_F(t)$, where $V_F(t)$ is in form (\ref{costfun}). It can be observed that $V_F(t)\leq e^{-0.1t}V_F(0)$ for all $t\geq0$, implying exponential convergence of the formation system. In conclusion, the simulation result illustrates Theorem \ref{th exponentially stable}. 

In fact, when we repeat the simulation by choosing other values of  $r$ in the same way as above, it can always be obtained that $V_F$ vanishes to zero exponentially and the target formation shape is eventually formed. Moreover, when we select each component of $r$ from the uniform distribution on $(-1,1)$, the target formation shape can still be formed in most cases. In other cases, the angle constraints can usually be satisfied with an exponentially fast speed, i.e., $V_F$ vanishes to zero exponentially, whereas the target formation shape is not eventually formed. This is because that $\mathcal{T}_\mathcal{G}^F$ is not sufficient for $(\mathcal{G},p^*)$ to be globally angle rigid. Note that the edge length of the pentagon formed by $q$ is $1.176$, therefore, the attraction region is sizable. In Fig. \ref{fig formation1}, the initial positions of agents are randomly set, it is shown that all the angle constraints are exponentially satisfied, but the agents form an incorrect shape.
\begin{figure}
	\centering
	\includegraphics[width=8.5cm]{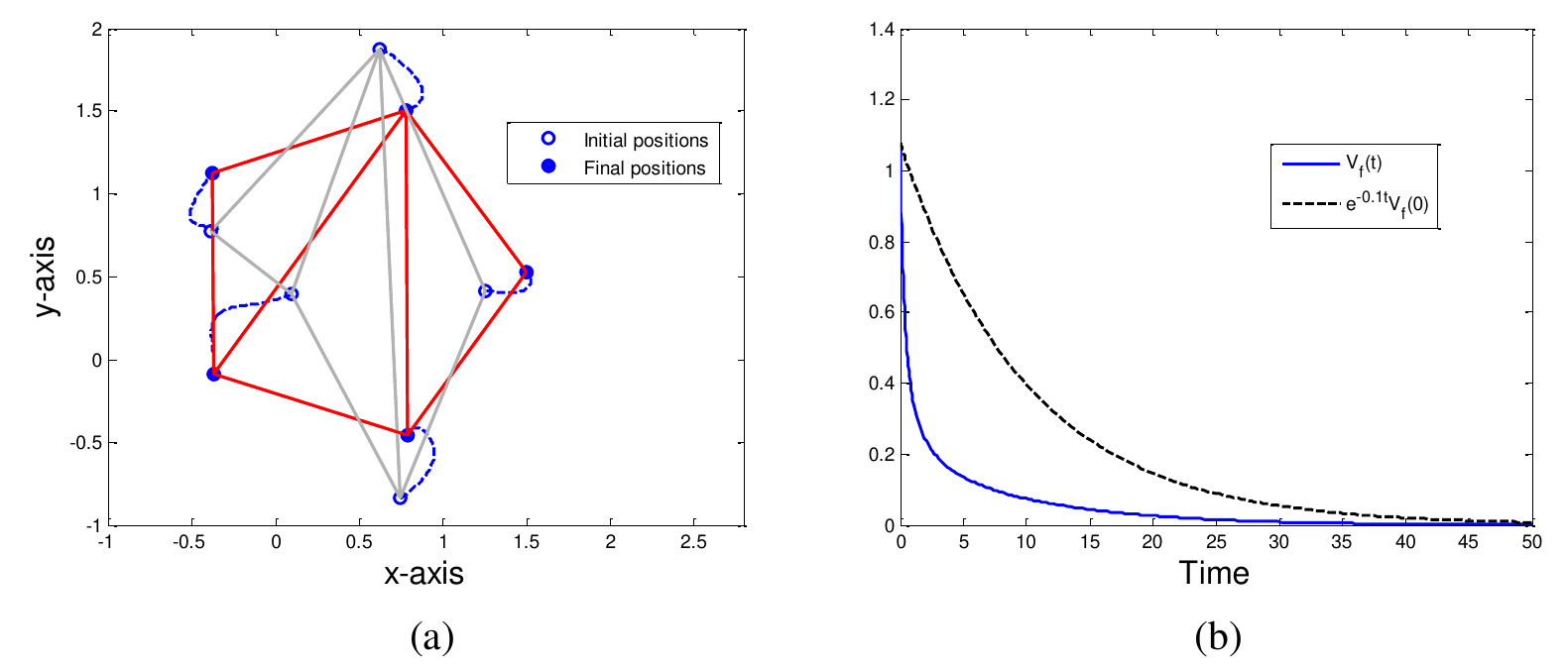}\\
	\caption{(a) Under control law (\ref{u_i}), the agents asymptotically form a regular pentagon. (b) $V_F(t)$ vanishes to zero in an exponential speed.}\label{fig formation}
\end{figure}
\begin{figure}
	\centering
	\includegraphics[width=8.5cm]{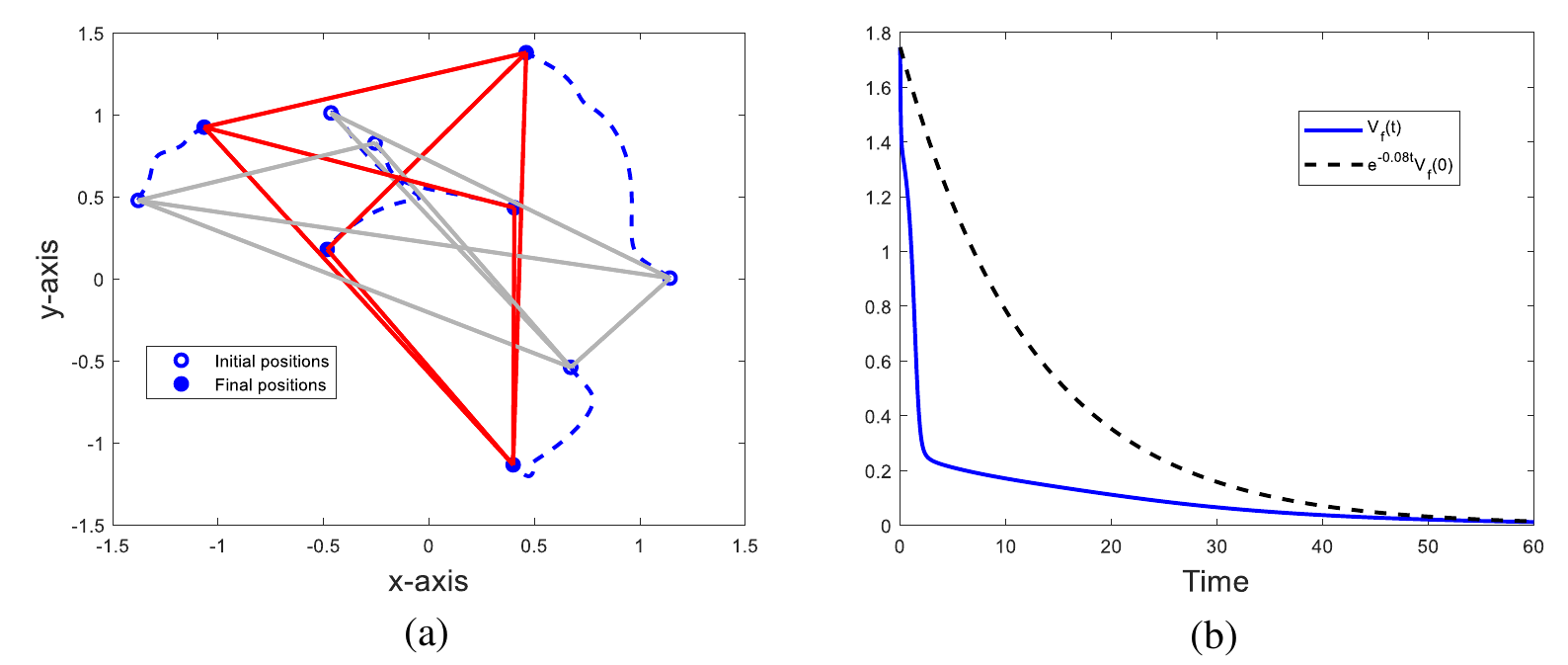}\\
	\caption{(a) Under control law (\ref{u_i}), the agents with a set of random initial positions asymptotically form a shape distinct to regular pentagon. (b) $V_F(t)$ vanishes to zero in an exponential speed.}\label{fig formation1}
\end{figure}
\end{exmp}
\begin{exmp}\label{example2}
Consider the framework in Fig. \ref{fig two triangulated graphs} (b) as the target formation. According to (\ref{T_G^f}), the set of desired angle information should be $\{g_{13}^{*T}g_{14}^*=0.8090, g_{14}^{*T}g_{15}^*=0.8090, g_{31}^{*T}g_{34}^*=0.3090, g_{41}^{*T}g_{45}^*=0.8090\}$. Under the same initial condition as in Example \ref{example1}, although $V_F$ vanishes to zero exponentially, the control law (\ref{u_i}) cannot stabilize the target formation, as shown in Fig. \ref{fig formation2}. This is because the angle constraints determined by $\mathcal{T}_\mathcal{G}^F$ are not sufficient to determine angle rigidity of the framework.
\begin{figure}
	\centering
	\includegraphics[width=8.5cm]{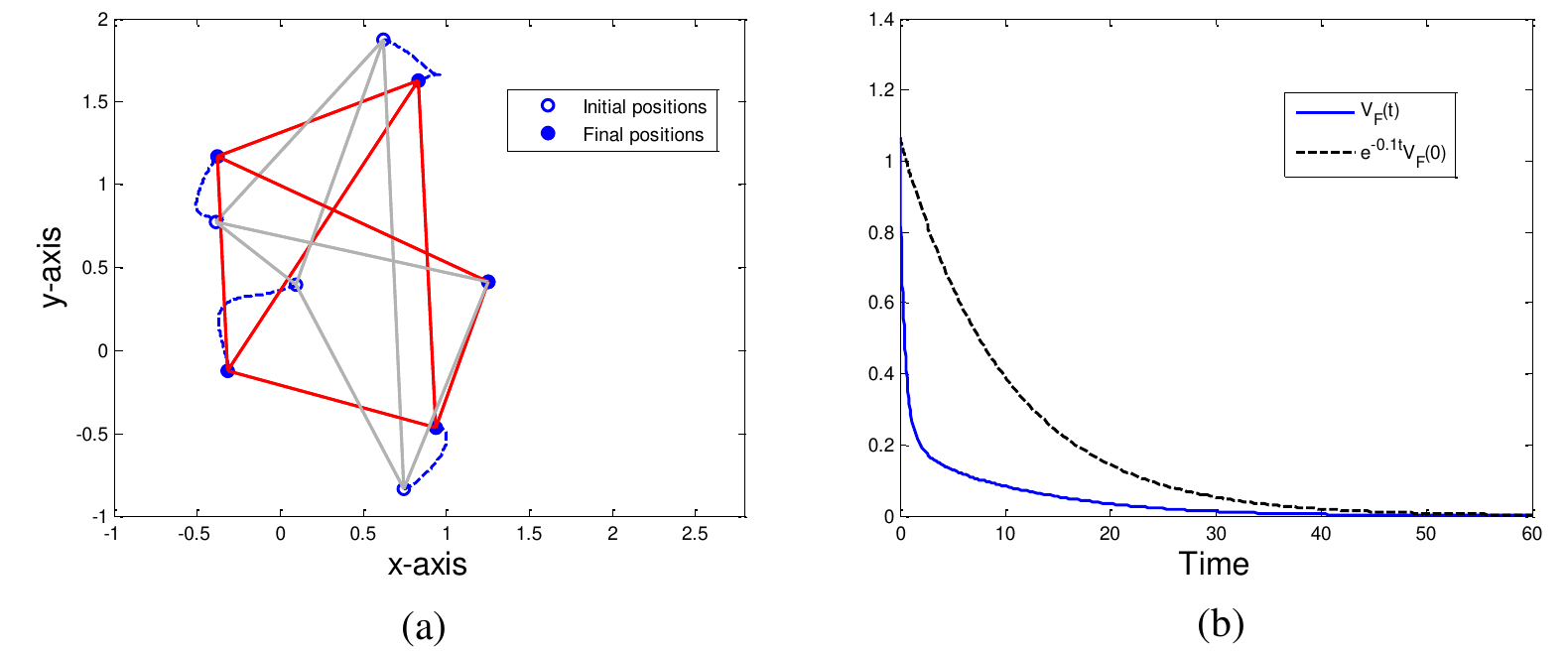}\\
	\caption{(a) Under control law (\ref{u_i}), the agents asymptotically form a shape distinct to a regular pentagon. (b) $V_F(t)$ vanishes to zero in an exponential speed.}\label{fig formation2}
\end{figure}
\end{exmp}
\begin{exmp}
In this example, we control orientation and scale of the formation formed in Example \ref{example1} by implementing the control input (\ref{u_i^M}). Let agents $3$ and $4$ be the two leaders. Now we aim to drive the direction of $p_3-p_4$ to be horizontal with respect to the global coordinate system, while setting the length of each edge as $0.5$. It suffices to set the target displacement between two leaders as $\tilde{p}_3-\tilde{p}_4=(-0.5,0)^T$. Fig. \ref{fig maneuver} shows the trajectories of agents and the evolution of $V(t)$ in (\ref{maneuver costfun}), in which we can observe the validity of Theorem \ref{th stability of maneuver control}. 
\end{exmp}

\section{Conclusion}\label{sec conclusion}

In this paper, we have developed an angle rigidity theory to study when a framework in the plane can be determined by angles uniquely up to translations, rotations, scalings and reflections. We have also proved that the shape of a triangulated framework can always be uniquely determined by angles. On the basis of the proposed angle rigidity theory, a distributed formation controller has been designed for formation shape stabilization. We have proved that by implementing our control strategy, a formation containing a strongly nondegenerate triangulated framework is locally exponentially stable. Taking the advantage of high degrees of freedom, we have proposed a distributed control strategy, which can drive agents to stabilize a target formation shape with prescribed orientation and scale. 

The angle rigidity theory proposed in this paper is only for graphs in the plane, similar definitions can be easily extended to higher dimensional spaces, but many properties of angle rigidity may become invalid. This is because many theoretical tools we have used cannot be directly applied to the higher dimensional case. We leave the angle rigidity theory in higher dimensional spaces as the future work. Moreover, the controller we presented requires agents to sense relative position states. We will try to design a bearing-only control law in future. 

% OR

%\begin{figure}
%\begin{center}
%\epsfig{file=jcaesar,width=7cm}
%\caption{Gaius Julius Caesar, 100--44 B.C.}
%\label{fig1}
%\end{center}
%\end{figure}

\bibliographystyle{plain}        % Include this if you use bibtex 
\bibliography{autosam}           % and a bib file to produce the 
                                 % bibliography (preferred). The
                                 % correct style is generated by
                                 % Elsevier at the time of printing.

\section{Appendix. A: Proof of Theorem \ref{th IAR=IBR}}\label{app A}

Necessity. Since $\Null(R_{\mathcal{T}_{\mathcal{G}}^*})=(\Null(R_g)\cap \range(R_B))\cup \Null(R_B)$, $\dim(\Null(R_{\mathcal{T}_{\mathcal{G}}^*}))$ reaches its minimum only if $\dim(\Null(R_B))$ is minimal. Recall that it always holds that $\Null(R_B)\supseteq\mathcal{S}_s\cup\mathcal{S}_t$, once  $(\mathcal{G},p)$ is infinitesimally angle rigid, it must hold that $\Null(R_B(p))=\mathcal{S}_s\cup\mathcal{S}_t$. That is, $(\mathcal{G},p)$ is infinitesimally bearing rigid.

Sufficiency. Note that $R_{\mathcal{T}_{\mathcal{G}}^*}=R_gR_B$, and infinitesimal bearing rigidity implies $\Null(R_B)=\mathcal{S}_t\cup\mathcal{S}_s$. To show $\Null(R_{\mathcal{T}_{\mathcal{G}}^*})=\mathcal{S}$, it suffices to show that for any $\eta\in \Null(R_g)\cap \range(R_B)$, we have $\eta=R_Bq$ for some $q\in\mathcal{S}_r$. 

Suppose $\eta=R_Bv$ and $R_g\eta=R_gR_Bv=0$ for some $v=(v_1^T,\cdots,v_n^T)^T\in\mathbb{R}^{2n}$. Let $g_{ij}^Tg_{ik}$ be a component of $f_{\mathcal{G}}$ such that $g_{ij}$ and $g_{ik}$ are not collinear. Then $R_{\mathcal{T}_{\mathcal{G}}^*}v=0$ implies that $\frac{\partial g_{ij}^Tg_{ik}}{\partial g}\diag(\frac{P_{ij}}{||e_{ij}||})\bar{H}v=0$, which is equivalent to
\begin{equation}\label{sum is zero}
e_{ik}^TP_{ij}(v_i-v_j)+e_{ij}^TP_{ik}(v_i-v_k)=0.
\end{equation}
Note that for any nonzero vectors $x,y\in\mathbb{R}^2$, $P(x)y$ is perpendicular to $x$. Therefore, there always exist $c_{ij},c_{ik}\in\mathbb{R}$ such that
\begin{equation}\label{two equalities}
P_{ij}(v_i-v_j)=c_{ij}\mathscr{R}_o(\frac{\pi}{2})g_{ij}, ~P_{ik}(v_i-v_k)=c_{ik}\mathscr{R}_o(\frac{\pi}{2})g_{ik}.
\end{equation}
It follows that
\begin{equation}\label{two modified equalities}
v_i-v_j=c_{ij}\mathscr{R}_o(\frac{\pi}{2})g_{ij}+c_{ij}'g_{ij}, ~v_i-v_k=c_{ik}\mathscr{R}_o(\frac{\pi}{2})g_{ik}+c_{ik}'g_{ik}
\end{equation}
for some $c_{ij}',c_{ik}'\in\mathbb{R}$. Substituting (\ref{two equalities}) into (\ref{sum is zero}), we have $$c_{ij}e_{ik}^T\mathscr{R}_o(\frac{\pi}{2})g_{ij}+ c_{ik}e_{ij}^T\mathscr{R}_o(\frac{\pi}{2})g_{ik}=0.$$
Note also that $\mathscr{R}_o^T(\frac{\pi}{2})=-\mathscr{R}_o(\frac{\pi}{2})$, then we have
$$(c_{ij}||e_{ik}||-c_{ik}||e_{ij}||)g_{ij}^T\mathscr{R}_o(\frac{\pi}{2})g_{ik}=0.$$
Since $g_{ij}$ and $g_{ik}$ are not collinear, $g_{ij}^T\mathscr{R}_o(\frac{\pi}{2})g_{ik}\neq0$. It follows that $c_{ij}||e_{ik}||=c_{ik}||e_{ij}||$. That is, $c_{ij}=c_{ijk}||e_{ij}||$, $c_{ik}=c_{ijk}||e_{ik}||$ for some $c_{ijk}\in\mathbb{R}$. Together with (\ref{two modified equalities}), we have
\begin{equation}\label{two final equalities}
\begin{split}
&v_i-v_j=c_{ijk}\mathscr{R}_o(\frac{\pi}{2})e_{ij}+\bar{c}_{ij}e_{ij},\\ &v_i-v_k=c_{ijk}\mathscr{R}_o(\frac{\pi}{2})e_{ik}+\bar{c}_{ik}e_{ik},
\end{split}
\end{equation}
where $\bar{c}_{ij}=c_{ij}'/||e_{ij}||$, $\bar{c}_{ik}=c_{ik}'/||e_{ik}||$.

So far we have proved that if $(i,j,k)\in\mathcal{T}_\mathcal{G}^*$ and $g_{ij}$ is not collinear with $g_{ik}$, then (\ref{two final equalities}) holds for some $c_{ijk}\in\mathbb{R}$. In the following, by constructing a $\mathcal{T}_{\mathcal{G}}^*$, we will show that there exists a common constant $c\in\mathbb{R}$ such that $v_i-v_j=c\mathscr{R}_o(\frac{\pi}{2})e_{ij}+\bar{c}_{ij}e_{ij}$ for all $(i,j)\in\mathcal{E}$.

Now we construct a set $\mathcal{T}_\mathcal{G}^*\subseteq\mathcal{T}_\mathcal{G}$ such that $g_{ij}$ and $g_{ik}$ are not collinear for all $(i,j,k)\in\mathcal{T}_\mathcal{G}^*$. Since $(\mathcal{G},p)$ is infinitesimally bearing rigid, from Lemma \ref{le IBR=IDR} and Lemma \ref{le noncollinear}, for any vertex $i$, there exist at least two neighbors $j,k\in\mathcal{N}_i$ such that $g_{ij}$ and $g_{ik}$ are not collinear. As a result, we can divide $\mathcal{N}_i$ into two sets $\hat{\mathcal{N}}_i$ and $\check{\mathcal{N}}_i$, such that for any $j\in\hat{\mathcal{N}}_i$ and $k\in\check{\mathcal{N}}_i$, $g_{ij}$ and $g_{ik}$ are not collinear. We construct a set $^i\mathcal{T}_{\mathcal{G}}^*$ by the following two steps:

Step 1. Select a vertex $j_1\in\hat{\mathcal{N}}_i$ randomly, let $(i,j_1,k)$(if $j_1<k$) or $(i,k,j_1)$(if $j_1>k$) for all $k\in\check{\mathcal{N}}_i$ be an element of $^i\mathcal{T}_{\mathcal{G}}^*$.

Step 2. Select a vertex $k_1\in\check{\mathcal{N}}_i$ randomly, let $(i,j,k_1)$(if $j<k_1$) or $(i,k_1,j)$(if $j>k_1$) for all $j\in\hat{\mathcal{N}}_i\setminus\{j_1\}$ be an element of $^i\mathcal{T}_{\mathcal{G}}^*$.

Let $\mathcal{T}_{\mathcal{G}}^*=\cup_{i\in\mathcal{V}}~^i\mathcal{T}_{\mathcal{G}}^*$. It is obvious that for any $i,j,k\in\mathcal{T}_{\mathcal{G}}^*$, $g_{ij}$ and $g_{ik}$ are not collinear. Now we regard each edge $(i,j)$ of $\mathcal{G}$ as a vertex of $\mathcal{G}'$, $(i,j)$ and $(i,k)$ are adjacent if $(i,j,k)$ or $(i,k,j)$ belongs to $\mathcal{T}_{\mathcal{G}}^*$. By our approach for construction of $^i\mathcal{T}_{\mathcal{G}}^*$, it is easy to see that for any $i\in\mathcal{V}$ and $j,k\in\mathcal{N}_i$, $(i,j)$ and $(i,k)$ are either adjacent or both neighbors of $(i,j_1)$ or $(i,k_1)$. Therefore, the graph $\mathcal{G}'$ corresponding to $\mathcal{T}_{\mathcal{G}}^*$ is connected. We regard $c_{ij}$ as the state corresponding to $(i,j)$ if $v_i-v_j=c_{ij}\mathscr{R}_o(\frac{\pi}{2})e_{ij}+\bar{c}_{ij}e_{ij}$. Note that (\ref{two final equalities}) implies that if $(i,j)$ and $(i,k)$ are adjacent, they share a common state $c_{ijk}\in\mathbb{R}$. Since $\mathcal{G}'$ is connected, all edges in $\mathcal{G}'$ have a consensus state $c\in\mathbb{R}$. That is, $v_i-v_j=c\mathscr{R}_o(\frac{\pi}{2})e_{ij}+\bar{c}_{ij}e_{ij}$ for all $(i,j)\in\mathcal{E}$.

This implies that $\bar{H}v=c(I_m\otimes\mathscr{R}_o(\frac{\pi}{2}))\bar{H}p+\bar{C}\bar{H}p$, where $\bar{C}=\diag(\bar{c}_{ij})\otimes I_2$. Then
\[
\begin{split}
\eta&=R_Bv=\diag(\frac{P_{ij}}{||e_{ij}||})\bar{H}v\\
&=\diag(\frac{P_{ij}}{||e_{ij}||})c(I_m\otimes\mathscr{R}_o(\frac{\pi}{2}))(H\otimes I_2)p+\diag(\frac{P_{ij}}{||e_{ij}||})\bar{C}\bar{H}p\\
&=\diag(\frac{P_{ij}}{||e_{ij}||})(H\otimes I_2)c(I_n\otimes\mathscr{R}_o(\frac{\pi}{2}))p \\
&=R_Bc(I_n\otimes\mathscr{R}_o(\frac{\pi}{2}))p.
\end{split}
\]
Since $c(I_n\otimes\mathscr{R}_o(\frac{\pi}{2}))p\in\mathcal{S}_r$, the proof is completed.

\section{Appendix. B: Proof of Theorem \ref{th angle congruence=bearing congruence}}\label{app B}

We first present some lemmas that are required to prove Theorem \ref{th angle congruence=bearing congruence}.

In \cite{Dongarra79}, the authors showed that for a positive semi-definite matrix $A\in\mathbb{R}^{n\times n}$ with $\rank(A)=r$, if $\Pi^TA\Pi=R^TR$ for a specified permutation matrix $\Pi\in\mathbb{R}^{n\times n}$, where $R\in\mathbb{R}^{r\times n}$, then this Cholesky decomposition is unique. Here the uniqueness of Cholesky decomposition implies that if $\bar{R}^T\bar{R}=\Pi^TA\Pi$ for some $\bar{R}\in\mathbb{R}^{r\times n}$, then $R=\mathscr{R}\bar{R}$ for some $\mathscr{R}\in \textrm{O(2)}$. It is straightforward to obtain the following lemma.

\begin{lem}\label{le unique cholesky decomposition}
	For a matrix $R\in\mathbb{R}^{r\times n}$ with $\rank(R)=r$, if $R^TR=\bar{R}^T\bar{R}$ for some $\bar{R}\in\mathbb{R}^{r\times n}$, then $R=\mathscr{R}\bar{R}$ for some  $\mathscr{R}\in O(r)$.
\end{lem}

Let $\mathscr{H}_{x}=\mathscr{H}(x)\triangleq I_2-2xx^T$ be a Householder transformation, here $x\in\mathbb{R}^2$ is a unit vector. Geometrically, $\mathscr{H}_xy$ with $y\in\mathbb{R}^2$ is a reflection of $y$ about the vector which is perpendicular to $x$. We list some properties of $\mathscr{H}_{x}$ in the following lemma.

\begin{lem}\label{le H_x}
	For any given unit vectors $x,y\in\mathbb{R}^2$, $\mathscr{H}_x$ has the following properties:
	
	(i) $\mathscr{H}_x^T=\mathscr{H}$, $\mathscr{H}_x^2=I_2$;
	
	(ii) $\mathscr{H}_x=\mathscr{R}_e(\theta)$ for some $\theta\in[0,2\pi)$;
	
	(iii) For any $\theta\in[0,2\pi)$, there exists a unit vector $z\in\mathbb{R}^2$ such that $\mathscr{H}_z=\mathscr{R}_e(\theta)$;
	
	(iv) The eigenspace of $\mathscr{H}_x$ associated with the eigenvalue 1 is $\Span\{x^{\perp}\}$.
\end{lem}

\textbf{Proof.} The statements in (i) and (ii) are easy to verify, thus the proofs are omitted here. Now we prove the rest of statements.

(iii) Let $x=(x_1,x_2)^T$, then $H_x=I-2xx^T=\begin{pmatrix}
1-2x_1^2~~ & -2x_1x_2 \\
-2x_1x_2~~ & 1-2x_2^2
\end{pmatrix}$. It suffices to show that $1-2x_1^2=-(1-2x_2^2)$ and $(1-2x_1^2)^2+(-2x_1x_2)^2=1$. Since $x_1^2+x_2^2=1$, the first equality obviously holds. For the second equality, we have
\[
\begin{split}
(1-2x_1^2)^2+(-2x_1x_2)^2&=1-4x_1^2+4x_1^4+4x_1^2x_2^2\\
&=1-4x_1^2+4x_1^2(x_1^2+x_2^2)=1.
\end{split}
\]

(iv) From the proof for (iii), it suffices to find suitable $x$ such that $1-2x_1^2=\cos\theta$, $1-2x_2^2=-\cos\theta$, and $-2x_1x_2=\sin\theta$. If $\theta\in[0,\pi)$, we can obtain a solution as $x_1=\sqrt{\frac{1-\cos\theta}{2}}$, $x_2=-\sqrt{\frac{1-\cos\theta}{2}}$. If $\theta\in[\pi,2\pi)$, a solution is $x_1=\sqrt{\frac{1-\cos\theta}{2}}$, $x_2=\sqrt{\frac{1-\cos\theta}{2}}$.

(v) Let $y\in\mathbb{R}^2$ be a vector such that $H_xy=y$, then $y-2x^Tyx=y$, which holds if and only if $x^Ty=0$, i.e., $y=c x^{\perp}$ for some constant $c\in\mathbb{R}$.
\QEDA

With the aid of Lemma \ref{le H_x}, we can establish the following result.

\begin{lem}\label{le A=BH}
	If $A\eta=B\eta$, where $A,B\in \textrm{O(2)}$, and $\eta\in\mathbb{R}^2$ is a unit vector, then $A=B$ or $A=B\mathscr{H}_{\eta^{\perp}}$.
\end{lem}

\textbf{Proof.} Note that a 2-dimensional orthogonal matrix is either a rotation matrix or a reflection matrix. Without loss of generality, we discuss the problem in three cases:

Case 1. $A=\mathscr{R}_o(\alpha)$, $B=\mathscr{R}_o(\beta)$ for some $\alpha, \beta\in[0,2\pi)$. Then $\mathscr{R}_o(\alpha)\eta=\mathscr{R}_o(\beta)\eta$, implying $\mathscr{R}_o(\alpha-\beta)\eta=\eta$. Hence $\alpha-\beta=0$. That is, $A=B$.

Case 2. $A=\mathscr{R}_e(\alpha)$, $B=\mathscr{R}_e(\beta)$ for some $\alpha, \beta\in[0,2\pi)$. Following the same procedure in Case 1, one can also obtain $A=B$.

Case 3. $A=\mathscr{R}_o(\alpha)$, $B=\mathscr{R}_e(\beta)=\mathscr{R}_o(\beta)\bar{I}$ for some $\alpha, \beta\in[0,2\pi)$. Then $\mathscr{R}_o(\alpha)\eta=\mathscr{R}_o(\beta)\bar{I}\eta$. It follows that $\eta=\mathscr{R}_o(\beta-\alpha)\bar{I}\eta=\mathscr{R}_e(\beta-\alpha)\eta$. From Lemma \ref{le H_x} (iii), there exists some $x\in\mathbb{R}^2$ such that $\mathscr{H}_x=\mathscr{R}_e(\beta-\alpha)$. That is, $\eta=\mathscr{H}_x\eta$. Using (iv) in Lemma \ref{le H_x}, we have $\eta\in \Span\{x^{\perp}\}$. Then $x=\pm\eta^{\perp}$, $\mathscr{H}_x=\mathscr{H}_{\eta^\perp}$. As a result, $\mathscr{H}_{\eta^\perp}=\mathscr{R}_e(\beta-\alpha)=\mathscr{R}_o(\beta-\alpha)\bar{I}$, implying that $\mathscr{R}_o(\alpha)\mathscr{H}_{\eta^\perp}=\mathscr{R}_e(\beta)$. By (i) in Lemma \ref{le H_x}, we have $A=\mathscr{R}_o(\alpha)=\mathscr{R}_e(\beta)\mathscr{H}_{\eta^\perp}=B\mathscr{H}_{\eta^\perp}$.
\QEDA

Let $\mathcal{F}$ denote a graph with $4$ vertices and 5 edges, then the following lemma holds.

\begin{lem}\label{le L_4}
	$(\mathcal{F},p)$ is infinitesimally bearing rigid if and only if $p$ is nondegenerate.
\end{lem}
The necessity of Lemma \ref{le L_4} is obvious. For sufficiency, it is easy to see that $\mathcal{F}$ must be a triangulated Laman graph. Since $(\mathcal{L}_4,p)$ is strongly nondegenerate if and only if $p$ is nondegenerate, from Lemma \ref{le strong nondegeneracy to idr}, 
$(\mathcal{L}_4,p)$ is infinitesimally distance rigid. 

With Lemmas \ref{le unique cholesky decomposition}, \ref{le H_x}, \ref{le A=BH} and \ref{le L_4} at hand, we now give the proof for Theorem \ref{th angle congruence=bearing congruence}.

\textbf{Proof of Theorem \ref{th angle congruence=bearing congruence}.}
We first note that $(I_n\otimes\mathscr{R})^{-1}q\in ~B_{\mathcal{K}}^{-1}(B_{\mathcal{K}}(p))$ is equivalent to $B_{\mathcal{K}}((I_n\otimes\mathscr{R})^{-1}q)=~B_{\mathcal{K}}(p)$, which is also equivalent to $B_{\mathcal{K}}(q)=(I_m\otimes\mathscr{R})B_{\mathcal{K}}(p)$. Therefore, it suffices to show that $f_\mathcal{K}(q)=f_\mathcal{K}(p)$ if and only if $B_{\mathcal{K}}(q)=(I_m\otimes\mathscr{R})B_{\mathcal{K}}(p)$.

Sufficiency. For any $i,j,k\in\mathcal{V}$, it is straightforward that $$g_{ij}^T(q)g_{ik}(q)=g_{ij}^T(p)\mathscr{R}^T\mathscr{R}g_{ik}(p) =g_{ij}^T(p)g_{ik}(p).$$

To prove necessity, we consider the following two cases.

Case 1. The configuration $p$ is degenerate. Let $\tilde{g}$ be a unit vector such that $\tilde{g}$ is collinear with $g_{ij}(p)$ for all $i,j\in\mathcal{V}$, then $g_{ij}(q)=\mathscr{R}g_{ij}(p)$ if and only if $g_{ij}(q)=\mathscr{R}\mathscr{H}_{\tilde{g}^{\perp}(p)}g_{ij}(p)$. For any $i,j\in\mathcal{V}$, let $\mathscr{R}_{ij}\in \textrm{O(2)}$ such that $g_{ij}(q)=\mathscr{R}_{ij}g_{ij}(p)$. To prove necessity, it suffices to show that for any distinct vertices $i,j,k\in\mathcal{V}$, if $g_{ij}(q)=\mathscr{R}_{ij}g_{ij}(p)$ and $g_{ik}(q)=\mathscr{R}_{ik}g_{ik}(p)$, there always holds  $\mathscr{R}_{ij}=\mathscr{R}_{ik}$ or $\mathscr{R}_{ij}=\mathscr{R}_{ik}\mathscr{H}_{\tilde{g}^{\perp}(p)}$. Without loss of generality, suppose $g_{ij}(p)=g_{ik}(p)$. Then $g_{ij}^T(q)g_{ik}(q)=g_{ij}^T(p)g_{ik}(p)=1$, which holds if and only if $g_{ij}(q)=g_{ik}(q)$, i.e., $\mathscr{R}_{ij}g_{ij}(p)=\mathscr{R}_{ik}g_{ik}(p)=\mathscr{R}_{ik}g_{ij}(p)$. By Lemma \ref{le A=BH}, $\mathscr{R}_{ij}=\mathscr{R}_{ik}$ or $\mathscr{R}_{ij}=\mathscr{R}_{ik}\mathscr{H}_{g_{ij}^{\perp}(p)}$. Since $\mathscr{H}_{g_{ij}^{\perp}(p)}=\mathscr{H}_{\tilde{g}^{\perp}(p)}$, the proof is completed.

Case 2. The configuration $p$ is nondegenerate. Note that $\mathcal{K}$ is complete, hence each vertex $i$ has at least two neighbors $j$ and $k$ such that $g_{ij}(p)$ and $g_{ik}(p)$ are not collinear. Then we can divide $\mathcal{N}_i$ into two sets $\hat{\mathcal{N}}_i$ and $\check{\mathcal{N}}_i$, such that for any $j\in\hat{\mathcal{N}}_i$ and $k\in\check{\mathcal{N}}_i$, $g_{ij}(p)$ and $g_{ik}(p)$ are not collinear. We first show that given $i\in\mathcal{V}$, for any $j\in\hat{\mathcal{N}}_i$, $k\in\check{\mathcal{N}}_i$, $l\in\mathcal{V}\setminus\{i,j,k\}$, it always holds that $G_{ijkl}(q)=\mathscr{R}_{ijkl}G_{ijkl}(p)$ for some $\mathscr{R}_{ijkl}\in \textrm{O(2)}$, where $G_{ijkl}=(g_{ij},g_{ik},g_{il},g_{jk},g_{jl},g_{kl})\in\mathbb{R}^{2\times6}$.

Since $\mathcal{K}$ is complete, we have $l\in\mathcal{N}_i$. Without loss of generality, we consider $l\in\check{\mathcal{N}}_i$. For the triangle composed of $i,j,k$, let $G_{ijk}=(g_{ij},g_{ik},g_{jk})\in\mathbb{R}^{2\times3}$. Since $f_{\mathcal{K}}(q)=~f_{\mathcal{K}}(p)$, we have $G_{ijk}^T(q)G_{ijk}(q)=G_{ijk}^T(p)G_{ijk}(p)$. Note that $g_{ij}(p)$ and $g_{ik}(p)$ are not collinear, thus we have $\rank(G_{ijk}(p))=2$. By virtue of Lemma \ref{le unique cholesky decomposition}, the Cholesky decomposition of $G_{ijk}^T(p)G_{ijk}(p)$ determines $G_{ijk}(p)$ up to a $2\times2$ orthogonal matrix $\mathscr{R}_{ijk}$. That is, $G_{ijk}(q)=\mathscr{R}_{ijk}G_{ijk}(p)$. Similarly, we have $G_{ijl}(q)=\mathscr{R}_{ijl}G_{ijl}(p)$ for $\mathscr{R}_{ijl}\in \textrm{O(2)}$. For vertices $j,k,l$, it follows from Case 1 that $G_{jkl}(q)=\mathscr{R}_{jkl}G_{jkl}(p)$ for $\mathscr{R}_{jkl}\in \textrm{O(2)}$ no matter $j,k,l$ are collinear or not. Since $\mathscr{R}_{ijk}g_{ij}(p)=\mathscr{R}_{ijl}g_{ij}(p)=g_{ij}(q)$, according to Lemma \ref{le A=BH}, $\mathscr{R}_{ijk}=\mathscr{R}_{ijl}$ or $\mathscr{R}_{ijk}=\mathscr{R}_{ijl}\mathscr{H}_{g_{ij}^{\perp}(p)}$.

Suppose that $\mathscr{R}_{ijk}\neq\mathscr{R}_{ijl}$, then
\[
\begin{split}
g_{jk}^T(p)g_{jl}(p)&=g_{jk}^T(q)g_{jl}(q)\\
&=g^T_{jk}(p)\mathscr{R}_{ijk}^T\mathscr{R}_{ijl}g_{jl}(p)\\
&=g^T_{jk}(p)\mathscr{H}_{g_{ij}^{\perp}(p)}g_{jl}(p)\\
&=g^T_{jk}(p)g_{jl}(p)-2g^T_{jk}(p)g_{ij}^{\perp}(p)g_{ij}^{\perp T}(p)g_{jl}(p).
\end{split}
\]
This implies $g^T_{jk}(p)g_{ij}^{\perp}(p)g_{ij}^{\perp T}(p)g_{jl}(p)=0$. Since $g_{ij}(p)$ and $g_{ik}(p)$ are not collinear, $g_{ij}(p)$ and $g_{jk}(p)$ are also not collinear. Similarly, $g_{ij}(p)$ and $g_{jl}(p)$ are not collinear. Thus a contradiction arises. We then have $\mathscr{R}_{ijk}=\mathscr{R}_{ijl}\triangleq\mathscr{R}_{ijkl}$, which implies that $\bar{G}_{ijkl}(q)=\mathscr{R}_{ijkl}\bar{G}_{ijkl}(p)$, where $\bar{G}_{ijkl}=(g_{ij},g_{ik},g_{il},g_{jk},g_{jl})\in\mathbb{R}^{2\times5}$. Consider the framework $(\mathcal{F},q)$, where $\mathcal{F}$ is a graph with vertex set $\{i,j,k,l\}$ and edge set $\{(i,j),(i,k),(i,l),(j,k),(k,l)\}$. Since these four vertices are not collinear, according to Lemma \ref{le L_4}, $(\mathcal{F},q)$ is infinitesimally bearing rigid, thus is globally bearing rigid. This implies that $g_{kl}(q)$ can be uniquely determined by $\bar{G}_{ijkl}(q)$. As a result, $G_{ijkl}(q)=\mathscr{R}_{ijkl}G_{ijkl}(p)$.

The above proof implies that given $i\in\mathcal{V}$, we have $$\mathscr{R}\triangleq\mathscr{R}_{ijk}=\mathscr{R}_{ijl}=\mathscr{R}_{ikl}=\mathscr{R}_{jkl}$$ for any $j\in\hat{\mathcal{N}}_i$, $k\in\check{\mathcal{N}}_i$, $l\in\mathcal{V}\setminus\{i,j,k\}$. Note that any edge in graph $\mathcal{K}$ is involved in a triangle including vertex $i$. Therefore, $g_{ij}(q)=\mathscr{R}g_{ij}(p)$ for any $(i,j)\in\mathcal{E}$.
\QEDA

\section{Appendix. C: Proof of Theorem \ref{th a unique shape}}\label{app C}

(i) From Lemma \ref{le strong nondegeneracy to idr} and the fact that $|\mathcal{E}_n|=2n-3$, strong nondegeneracy and minimal infinitesimal angle rigidity are equivalent for $(\mathcal{L}_n,p)$. Next we show $\mathcal{T}_{\mathcal{L}_n}^*$ in (\ref{T_L^*}) is minimally suitable for $(\mathcal{L}_n,p)$ to be minimally infinitesimally angle rigid. 

By virtues of Theorem \ref{th IAR=IBR} and Lemma \ref{le strong nondegeneracy to idr}, $(\mathcal{L}_n,p)$ is infinitesimally bearing rigid. Then $\Null(R_B)=\mathcal{S}_s\cup\mathcal{S}_t$. It suffices to show that for any $\eta\in \Null(R_g)\cap \range(R_B)$, there always exists $q\in\mathcal{S}_r$ such that $\eta=R_Bq$. Suppose that $R_{\mathcal{T}_{\mathcal{L}_n}^*}v=R_gR_Bv=0$ and $R_Bv\neq0$, where $v=(v_1^T,\cdots, v_n^T)\in\mathbb{R}^{2n}$. In the proof of Theorem \ref{th IAR=IBR}, we have shown that for any $(i,j,k)\in\mathcal{T}_{\mathcal{L}_n}^*$, if $g_{ij}$ is not collinear with $g_{ik}$, then (\ref{two final equalities}) holds for some $c_{ijk}$. Recall that $(\mathcal{L}_n,p)$ is strongly nondegenerate, then for any $(i,j,k)\in\mathcal{T}_{\mathcal{L}_n}^*$, (\ref{two final equalities}) holds for some $c_{ijk}$. Without loss of generality, suppose $i<j<k$. Due to the definition in (\ref{T_L^*}), for each triangle in $\mathcal{L}_n$ formed by vertices $i$, $j$ and $k$, we have $(i,j,k),(j,i,k)\in\mathcal{T}_{\mathcal{L}_n}^*$. Now we regard $(i,j)$ as a vertex of $\mathcal{G}'$ for all $(i,j)\in\mathcal{E}_n$, two vertices in $\mathcal{G}'$ are adjacent if they belong to a same triangle in $\mathcal{L}_n$. Let $c_{ij}$ be the state of $(i,j)$ if $v_i-v_j=c_{ij}\mathscr{R}_o(\frac{\pi}{2})e_{ij}+\bar{c}_{ij}e_{ij}$ for some $\bar{c}_{ij}\in\mathbb{R}$. It is easy to see that $(i,j)$, $(i,k)$ and $(k,j)$ have a common state, implying that adjacent vertices in $\mathcal{G}'$ must have a common state. Note that in every step during the generation of graph $\mathcal{L}_n$, a new triangle is generated based on an existing edge. Therefore, $\mathcal{G}'$ must be connected. As a result, there exists a constant $c\in\mathbb{R}$ such that $v_i-v_j=c\mathscr{R}_o(\frac{\pi}{2})e_{ij}+\bar{c}_{ij}e_{ij}$ for all $(i,j)\in\mathcal{E}_n$. By similar analysis to the proof of Theorem \ref{th IAR=IBR}, we can obtain $v\in\mathcal{S}_r$, which implies that $(\mathcal{L}_n,p)$ is infinitesimally angle rigid for $\mathcal{T}_{\mathcal{L}_n}^*$. Moreover, observe that $|\mathcal{T}_{\mathcal{L}_n}^*|=2n-4$, we conclude that $\mathcal{T}_{\mathcal{L}_n}^*$ is also minimal.

(ii). We prove the statement by induction.

For $n=3$, it is obvious that $(\mathcal{L}_3,p)$ with $\mathcal{T}_{\mathcal{L}_3}^\dagger=\mathcal{T}_{\mathcal{L}_3}^*=\{(1,2,3),(2,1,3)\}$ is globally angle rigid.

For $n\geq4$, suppose that $(\mathcal{L}_{n-1},p)$ with $\mathcal{T}_{\mathcal{L}_{n-1}}^\dagger$ is globally angle rigid. Next we show that $(\mathcal{L}_n,p)$ is globally angle rigid with $\mathcal{T}_{\mathcal{L}_{n}}^\dagger$. Without loss of generality, let $i$ and $j$ be the neighbors of $n$ and $i<j$. Note that $(i,j)\in\mathcal{E}_{n-1}$, and $i$, $j$ must have at least one common neighbor vertex in $\mathcal{L}_{n-1}$, let $k$ be the minimum index among them, it is easy to see $\mathcal{T}_{\mathcal{L}_n}^\dagger=\{(i,j,n),(j,i,n),(i,k,n)\}\cup\mathcal{T}_{\mathcal{L}_{n-1}}^\dagger$. It suffices to show that for any $q$ such that $f_{\mathcal{T}_{\mathcal{L}_n}^\dagger}(p)=f_{\mathcal{T}_{\mathcal{L}_n}^\dagger}(q)$, it always holds $f_{\mathcal{K}_n}(p)=f_{\mathcal{K}_n}(q)$. Since $\mathcal{L}_{n-1}$ is globally angle rigid, by Theorem \ref{th angle congruence=bearing congruence}, there exists a matrix $\mathscr{R}_{n-1}\in \textrm{O(2)}$ such that $g_{i'j'}(p)=\mathscr{R}_{n-1}g_{i'j'}(q)$ for all $i',j'\in\mathcal{V}_{n-1}$. From $g_{ij}^T(p)g_{in}(p)=g_{ij}^T(q)g_{in}(q)$, and $g_{ji}^T(p)g_{jn}(p)=g_{ji}^T(q)g_{jn}(q)$, we have $G_{ijn}^T(p)G_{ijn}(p)=G_{ijn}^T(q)G_{ijn}(q)$, where $G_{ijn}=(g_{ij},g_{nj},g_{ni})\in\mathbb{R}^{2\times3}$. Using strong nondegeneracy of $(\mathcal{L}_n,p)$, we have $\rank(G_{ijn}(p))=2$. By Lemma \ref{le unique cholesky decomposition}, $G_{ijn}(p)=\mathscr{R}_{ijn}G_{ijn}(q)$ for some $\mathscr{R}_{ijn}\in \textrm{O(2)}$. It follows that $g_{ij}(p)=\mathscr{R}_{n-1}g_{ij}(q)=\mathscr{R}_{ijn}g_{ij}(q)$. According to Lemma \ref{le A=BH}, $\mathscr{R}_{n-1}=\mathscr{R}_{ijn}$ or $\mathscr{R}_{n-1}=\mathscr{R}_{ijn}\mathscr{H}_{g_{ij}^{\perp}(q)}$.

Suppose that $\mathscr{R}_{n-1}\neq\mathscr{R}_{ijn}$, then $g_{ik}(p)=\mathscr{R}_{n-1}g_{ik}(q)=\mathscr{R}_{ijn}\mathscr{H}_{g_{ij}^{\perp}(q)}g_{ik}(q)$, and $g_{in}(p)=\mathscr{R}_{ijn}g_{in}(q)$. It follows that
\begin{equation}\label{gikgin}
\begin{split}
g_{ik}^T(p)g_{in}(p)&=g_{ik}^T(q)\mathscr{H}_{g_{ij}^{\perp}(q)}\mathscr{R}_{ijn}^T\mathscr{R}_{ijn}g_{in}(q)\\ &=g_{ik}^T(q)\mathscr{H}_{g_{ij}^{\perp}(q)}g_{in}(q).
\end{split}
\end{equation}
Recall that $(i,k,n)\in\mathcal{T}_{\mathcal{L}_n}^\dagger$, we have $g_{ik}^T(p)g_{in}(p)=g_{ik}^T(q)g_{in}(q)$. Together with (\ref{gikgin}), it follows that $g_{ik}^T(q)\mathscr{H}_{g_{ij}^{\perp}(q)}g_{in}(q)=g_{ik}^T(q)g_{in}(q)$, which holds if and only if $g_{ij}(q)$ is collinear with either $g_{ik}(q)$ or $g_{in}(q)$, i.e., either $(q_i^T,q_j^T,q_k^T)^T$ or $(q_i^T,q_j^T,q_n^T)^T$ is degenerate, implying that either $(p_i^T,p_j^T,p_k^T)^T$ or $(p_i^T,p_j^T,p_n^T)^T$ is degenerate. This conflicts with strong nondegeneracy of $(\mathcal{L}_n,p)$.  Therefore, $\mathscr{R}_{n-1}=\mathscr{R}_{ijn}\triangleq\mathscr{R}_{n}$. That is, $g_{i'j'}(p)=\mathscr{R}_{n}g_{i'j'}(q)$ for any $(i',j')\in\mathcal{E}_n$. It follows that $f_{\mathcal{K}}(p)=~f_{\mathcal{K}}(q)$. Hence, $(\mathcal{L}_{n},p)$ with $\mathcal{T}_{\mathcal{L}_{n}}^*$ is globally angle rigid.

\end{document}